\documentclass[journal,9pt,draftclsnofoot,onecolumn]{IEEEtran}
\usepackage{amsmath,epsfig}
\usepackage[lined,boxed,commentsnumbered]{algorithm2e}
\usepackage{amsmath,epsfig, citesort,cite}
\usepackage{algorithmic}
\usepackage{graphicx}
\usepackage{url}
\usepackage{amssymb,amsthm}
\usepackage{times,amsmath,amssymb,amsfonts,epsfig,graphicx}
\usepackage{enumerate}
\usepackage{stackrel}
\usepackage{multirow}
\usepackage{epsfig}
\usepackage{subfigure}
\usepackage{psfrag}
\usepackage{color}
\usepackage{comment}
\usepackage{framed}

\newcommand{\vectornorm}[1]{\|#1\|}
\newcommand{\vectornormbig}[1]{\big\|#1\big\|}

\newcommand{\obs}{\boldsymbol{y}}
\newcommand{\sensing}{\boldsymbol{\mathcal{A}}}

\newcommand{\signal}{\mathbf{X}}
\newcommand{\bestsignal}{\mathbf{X}^\ast}

\newcommand{\noise}{\boldsymbol{\varepsilon}}
\newcommand{\dimension}{m \times n}
\newcommand{\numsam}{p}
\newcommand{\sparsity}{s}
\newcommand{\id}{I}
\newcommand{\rank}{k}

\DeclareMathOperator*{\argmin}{arg\,min}

\newtheorem{lemma}{Lemma}

\newtheorem{theorem}{Theorem}
\newtheorem{corollary}{Corollary}

\title{MATRIX ALPS: Accelerated Low Rank and Sparse \\ Matrix Reconstruction}
\author{\IEEEauthorblockN{Anastasios Kyrillidis and Volkan Cevher \\}
\IEEEauthorblockA{Laboratory for Information and Inference Systems\\
Ecole Polytechnique Federale de Lausanne\\
\{anastasios.kyrillidis,volkan.cevher\}@epfl.ch}}
	
\begin{document}

\maketitle
\begin{abstract}
We propose \textsc{Matrix ALPS} for recovering a sparse plus low-rank decomposition of a matrix given its corrupted and incomplete linear measurements. Our approach is a first-order projected gradient method over non-convex sets, and it exploits a well-known memory-based acceleration technique. We theoretically characterize the convergence properties of \textsc{Matrix ALPS} using the stable embedding properties of the linear measurement operator. We then numerically illustrate that our algorithm outperforms the existing convex as well as non-convex state-of-the-art algorithms in computational efficiency without sacrificing stability. 
\end{abstract}

\section{Introduction}{\label{sec:prel}}  

Finding a low rank plus sparse matrix decomposition from a set of---possibly incomplete and noisy---measurements is critical in many applications. The list has expanded over the last ten years: examples include MRI signal processing, collaborative filtering, hyperspectral image analysis, large-scale data processing, etc. 
A general statement of the problem under consideration can be described as follows:

\textsc{Problem}. {\it Given a linear operator $\sensing: \mathbb{R}^{\dimension} \rightarrow \mathbb{R}^{\numsam} $ and a set of observations $\obs \in \mathbb{R}^{\numsam} $ (usually $\numsam \ll m \times n$):}  
\begin{align}
\obs = \sensing \bestsignal + \noise, \label{eq:observations}  
\end{align} {\it where $ \bestsignal := \mathbf{L}^{\ast} + \mathbf{M}^{\ast} \in \mathbb{R}^{\dimension} $ is the superposition of a rank-$\rank$ $\mathbf{L}^{\ast}$ and a $ \sparsity $-sparse $\mathbf{M}^{\ast} $ component that we desire to recover, identify a matrix $ \widehat{\mathbf{L}} \in \mathbb{R}^{\dimension} $ of rank (at most) $\rank$ and a matrix $\widehat{\mathbf{M}} \in \mathbb{R}^{\dimension} $ with sparsity level $\vectornormbig{\widehat{\mathbf{M}}}_0 \leq \sparsity $ such that:}   
\begin{align}
\big \lbrace \widehat{\mathbf{L}}, ~\widehat{\mathbf{M}} \big \rbrace = \argmin_{\substack{\mathbf{L},~\mathbf{M}: ~\text{rank}(\mathbf{L}) \leq \rank, ~\vectornorm{\mathbf{M}}_0 \leq \sparsity}}\vectornormbig{\obs - \sensing (\mathbf{L} + \mathbf{M})}_2. \label{eq:problem} \vspace{-1cm}
\end{align} Here, $ \noise \in \mathbb{R}^{\numsam} $ represents the potential noise term. For different linear operator $ \sensing $ and signal $\bestsignal$ configurations, the above problem arises in various research fields. Next, we briefly address some of the frameworks that (\ref{eq:problem}) is involved.  

\subsection{Compressed sensing and affine rank minimization}  

In the standard Compressed Sensing (CS) framework, we desire to reconstruct a $n$-dimensional, $ \sparsity $-sparse loading vector through a $\numsam$-dimensional set of observations with $\numsam \ll n$. This problem can be solved by finding the minimizer $\widehat{\mathbf{X}} := \widehat{\mathbf{M}} $ of:  
\begin{align}
\big \lbrace \widehat{\mathbf{M}} \big \rbrace = \argmin_{\substack{\mathbf{M}~: \mathbf{M} \in \mathbb{D}^n, ~\vectornorm{\mathbf{M}}_0 \leq \sparsity}}\vectornormbig{\obs - \sensing \mathbf{M}}_2. \label{eq:CS_problem}  
\end{align} where we reserve $ \mathbb{D}^n $ to denote the set of $ n \times n $ diagonal matrices. To establish solution uniqueness and reconstruction stability in (\ref{eq:CS_problem}), $ \sensing $ is usually assumed to satisfy the {\it sparse restricted isometry property} (sparse-RIP) \cite{candes06robust} where:  
\begin{align}
(1 - \delta_{\sparsity}(\sensing))\vectornormbig{\signal}_F \leq \vectornormbig{\sensing\signal}_2 \leq (1+\delta_{\sparsity}(\sensing))\vectornormbig{\signal}_F,  
\end{align} $ ~\forall \signal \in \mathbb{D}^{n}~~\text{with}~~\vectornormbig{\signal}_0 \leq\sparsity $ and $ \delta_{\sparsity}(\sensing) \in (0, 1) $. 

In the general affine rank minimization (ARM) problem, we aim to recover a low-rank matrix $ \bestsignal := \mathbf{L}^{\ast} $ from a set of observations $ \obs \in \mathbb{R}^{\numsam} $, according to (\ref{eq:observations}). 
The challenge is to reconstruct the true matrix given $ \numsam \ll m\cdot n $. A practical means to tackle this problem is by finding the simplest solution $ \widehat{\signal} := \widehat{\mathbf{L}} $ of minimum rank that minimizes the data error as:  
\begin{align}
\big \lbrace \widehat{\mathbf{L}} \big \rbrace = \argmin_{\substack{\mathbf{L}:~\text{rank}(\mathbf{L}) \leq \rank}}\vectornormbig{\obs - \sensing \mathbf{L}}_2. \label{eq:ARM_problem}  
\end{align} \cite{brecht2010} provides guarantees for exact and unique solution using the rank-RIP property for affine transformations where $ \sensing $ satisfies:
\begin{align}
(1 - \delta_{\rank}(\sensing))\vectornormbig{\signal}_F \leq \vectornormbig{\sensing\signal}_2 \leq (1+\delta_{\rank}(\sensing))\vectornormbig{\signal}_F,  
\end{align} $ ~\forall \signal \in \mathbb{R}^{\dimension}~~\text{with}~~\text{rank}(\signal) \leq\rank $ and $ \delta_{\rank}(\sensing) \in (0, 1) $.   
 
\subsection{Fusing low-dimensional embedding models}  
Robust Principal Component Analysis (RPCA) deals with the challenge of recovering a low rank and a sparse matrix component from a {\it complete} data matrix. In mathematical terms, we acquire a finite set of observations $ \mathbf{Y} \in \mathbb{R}^{\dimension} $ according to $\mathbf{Y} = \mathbf{L}^{\ast} + \mathbf{M}^{\ast}$
with $\mathbf{L}^{\ast} \in \mathbb{R}^{\dimension} $ and $\mathbf{M}^{\ast} \in \mathbb{R}^{\dimension}$, defined above. The ``robust'' characterization of the RPCA problem refers to $\mathbf{M}^{\ast}$ having {\it gross} non-zero entries with {\it arbitrary} energy. Under mild assumptions concerning the incoherence between $\mathbf{L}^{\ast}$ and $\mathbf{M}^{\ast}$ \cite{candes2009robust}, we can efficiently reconstruct both the low-rank and sparse components using convex and non-convex optimization approaches \cite{candes2009robust,zhou2011godec}.   

\subsection{Contributions}   
While solving the RPCA problem itself is a difficult task, here we assume: $(i)$ $\sensing$ is an arbitrary linear operator satisfying both sparse- and rank-RIP (this assumption includes the identity linear map of RPCA as a special case) and, $(ii)$ the total number of observations in $\obs $ is much less compared to the total number of variables we want to recover, i.e., $\numsam \ll m\cdot n$. Our contributions are two-fold:
\begin{itemize}
\item For noisy settings and arbitrary operator $\sensing $ satisfying sparse- and rank-RIP, we provide better restricted isometry constant guarantees compared to state-of-the-art approaches \cite{sparcs}.
\item We introduce \textsc{Matrix ALPS}, an accelerated, memory-based algorithm along with preliminary convergence analysis. 
\end{itemize}

The organization of the paper is as follows. In Section \ref{sec:ALPS}, we describe the algorithms in a nutshell and present the main theorem of the paper in Section \ref{section:convergence}. In Section \ref{sec:accelerated} we briefly study acceleration techniques in the recovery process. We provide empirical support for our claims for better data recovery performance and reduced complexity in Section \ref{section:experiments}. 

\textbf{Notation:} We reserve lower-case letters for scalar variable representation. Bold upper-case letters denote matrices while bold calligraphic upper-case letters represent linear maps. We reserve plain calligraphic upper-case letters for set representations. We denote a set of orthonormal, rank-1 matrices that span the subspace induced by $ \signal $ as $ \rm{ortho}(\signal) $. Given a matrix $ \signal $ and a subspace set $ \mathcal{S} $ such that $ \rm{span}(\mathcal{S}) \subseteq \rm{span}(\rm{ortho}(\signal)) $, the orthogonal projection of $ \signal $ onto the subspace spanned by $ \mathcal{S} $ is given by $ \mathcal{P}_{\mathcal{S}}\signal $ while $ \mathcal{P}_{\mathcal{S}}^{\bot}\signal $ represents the projection onto the subspace, orthogonal to $ \rm{span}(\mathcal{S}) $. Given a matrix $ \signal $ and an index set $ \mathcal{U} $, $ (\signal)_{\mathcal{U}} $ denotes the (sub)matrix of $ \signal $ with entries in $ \mathcal{U} $ while $ (\signal)_{\mathcal{U}^c} $ denotes the (sub)matrix of $ \signal $ with entries in the complement set of $ \mathcal{U} $. The best $ \sparsity $-sparse and rank-$ \rank $ approximations of a matrix $ \signal $ are given by $ \mathcal{P}_{\Sigma_{\sparsity}}(\signal) $ and $ \mathcal{P}_{\rank}(\signal) $, respectively. For any two subspace sets $ \mathcal{S}_1, ~\mathcal{S}_2 $, we use the shorthand $ \mathcal{P}_{\mathcal{S}_1 \setminus \mathcal{S}_2} $ to denote the projection onto the subspace defined by $ \mathcal{S}_1 $, orthogonal to the subspace defined by $ \mathcal{S}_2 $---similar notation is used for index sets. We use $ \signal_i \in \mathcal{R}^{\dimension}$ to represent the current matrix estimate at the $i$-th iteration. The rank of $ \signal $ is denoted as $ \text{rank}(\signal) \leq \min\lbrace m, n \rbrace $ while the non-zero index set of $ \signal $ is given by $ \rm{supp}(\signal) $. The empirical data error $f(\signal) := \vectornorm{\obs - \sensing \signal}_2^2 $ has gradient $ \nabla f(\signal) := -2 \sensing^\ast(\obs - \sensing \signal)$,  where $\sensing^\ast $ is the adjoint linear operator. $ \mathbb{I} $ represents the identity matrix.

\section{The \textsc{SpaRCS} Algorithm}{\label{sec:ALPS}} 

\begin{algorithm}[t]
   \caption{SpaRCS}\label{algo: class}
\begin{algorithmic}[1]
   \STATE {\bfseries Input:} $\obs$, $\sensing$, $ \sensing^{\ast} $,  Tolerance $ \eta $, MaxIterations
   \STATE {\bfseries Initialize:} $ \lbrace \mathbf{L}_0, \mathbf{M}_0 \rbrace \leftarrow 0 $, $ \lbrace \mathcal{L}_0, \mathcal{M}_0 \rbrace \leftarrow \lbrace \emptyset \rbrace $, $ i \leftarrow 0 $
   \STATE {\bfseries repeat} 
   \STATE \hspace{0.02cm}  $ \mathcal{S}_i^{\mathcal{L}} \leftarrow \mathcal{D}_i^{\mathcal{L}} \cup \mathcal{L}_i $ where $ \mathcal{D}_i^{\mathcal{L}} \leftarrow \rm{ortho}$ $\big(\mathcal{P}_{\rank}(\nabla f(\signal_{i}))\big) $
   \STATE \hspace{0.02cm} $ \mathcal{S}_i^{\mathcal{M}} \leftarrow \mathcal{D}_i^{\mathcal{M}} \cup \mathcal{M}_i $ where $ \mathcal{D}_i^{\mathcal{M}} \leftarrow \rm{supp}$ $\big(\mathcal{P}_{\Sigma_{\sparsity}}(\nabla f(\signal_{i}))\big) $
   
   \STATE \hspace{0.02cm} \textbf{Low rank matrix estimation:}
   \STATE \hspace{0.25cm} $ \mathbf{V}_{i}^{\mathcal{L}} \leftarrow \argmin_{\mathbf{V}: \mathbf{V} \in \rm{span}(\mathcal{S}_i^{\mathcal{L}})} \vectornormbig{\obs - \sensing (\mathbf{V} + \mathbf{M}_i)}_2^2 $ 
   \STATE \hspace{0.25cm} $ \mathbf{L}_{i+1} \leftarrow \mathcal{P}_{\rank}(\mathbf{V}_{i}^{\mathcal{L}}) ~~\text{with }~  \mathcal{L}_{i+1} \leftarrow \rm{ortho}$ $(\mathbf{L}_{i+1}) $  
   \STATE \hspace{0.02cm} \textbf{Sparse matrix estimation:}
   \STATE \hspace{0.25cm} $ \mathbf{V}_{i}^{\mathcal{M}} \leftarrow \argmin_{\mathbf{V}: \mathbf{V} \in \rm{supp}(\mathcal{S}_i^{\mathcal{M}})} \vectornormbig{\obs - \sensing (\mathbf{V} + \mathbf{L}_i)}_2^2 $ 
   \STATE \hspace{0.25cm} $ \mathbf{M}_{i+1} \leftarrow \mathcal{P}_{\Sigma_{\sparsity}}(\mathbf{V}_{i}^{\mathcal{M}}) ~~\text{with }~  \mathcal{M}_{i+1} \leftarrow \rm{supp}$ $(\mathbf{M}_{i}) $ 
   
   \STATE \hspace{0.02cm} $\signal_{i+1} \leftarrow \mathbf{L}_{i+1} + \mathbf{M}_{i+1}$
   \STATE \hspace{0.02cm} $ i \leftarrow i + 1 $
   \STATE {\bfseries until} $\vectornorm{\signal_{i} - \signal_{i-1}}_2 \leq \eta \vectornorm{\signal_{i}}_2 $ or MaxIterations.
\end{algorithmic}
\end{algorithm}

Explicit description of SpaRCS \cite{sparcs} is provided in Algorithm 1 in pseudocode form. This approach borrows from a series of vector and matrix reconstruction algorithms such as CoSaMP \cite{cosamp} and ADMiRA \cite{admira2010}. In a nutshell, this algorithm simply seeks to improve the current subspace and support set selection by iteratively collecting extended sets $ \mathcal{S}_i^{\mathcal{L}} $ and $ \mathcal{S}_i^{\mathcal{M}} $ with $ |\mathcal{S}_i^{\mathcal{L}}| \leq 2\rank $ and $ |\mathcal{S}_i^{\mathcal{M}}| \leq 2\sparsity $, respectively. Then, $ \sparsity $-sparse and rank-$ \rank $ matrices are estimated to fit the measurements in these restricted subspace/support sets using least squares techniques. 

\section{Improved Convergence Guarantees}{\label{section:convergence}}  
Before we present our analysis, we note the following. The reconstruction of both $\mathbf{L}^{\ast}$ and $\mathbf{M}^{\ast} $ from $\obs$ makes sense under mild conditions on $\mathbf{L}^{\ast}$ and $\mathbf{M}^{\ast}$. Borrowing from \cite{candes2009robust}, we assume that the low rank component $\mathbf{L}^{\ast}$ is not sparse and uniformly bounded with respect to its singuar vectors and the sparse compoment $\mathbf{M}^{\ast}$ is not low rank with support set uniformly random over the entries of $\mathbf{M}^{\ast}$. 

An important ingredient for our matrix analysis is the following lemma---the proof can be found in \cite{sparcs}.  
\begin{lemma}\label{lemma:1}
Let $ \mathcal{F} $ be a support set with $ |\mathcal{F}| \leq \sparsity $ and assume $ \mathbf{L} \in \mathbb{R}^{\dimension} $ is a rank-$ \rank $ matrix, satisfying the conditions above. Then, given a general linear operator $ \sensing: \mathbb{R}^{\dimension} \rightarrow \mathbb{R}^{\numsam} $ satisfying both sparse- and rank-RIP, we have:  
\begin{align}
\vectornormbig{(\sensing^{\ast}\sensing \mathbf{L})_{\mathcal{F}}}_F \lesssim \delta_{\sparsity + \rank}(\sensing)\vectornormbig{\mathbf{L}}_F, ~~\text{for $ \min\lbrace m, n \rbrace \gg \sparsity \gg \rank $}. \nonumber
\end{align}  where $\delta_{\sparsity + \rank}(\sensing)$ denotes the RIP constant of $\sensing$ over (disjoint) sparse index and low-rank subspace sets where the combined cardinality is less than $\sparsity + \rank$.
\end{lemma}

We provide improved conditions for convergence for Algorithm 1. The details of the proof can be found in the Appendix. The following theorem characterizes Algorithm 1:
\begin{theorem}\label{theorem:1}
Given the problem configuration described in (\ref{eq:observations}) and (\ref{eq:problem}), assume the linear operator $ \sensing $ satisfies the sparse-RIP and rank-RIP for $ \delta_{4\sparsity}(\sensing) \leq 0.075 $, $ \delta_{4\rank}(\sensing) \leq 0.04 $ and $\delta_{2\sparsity + 3\rank}(\sensing) \leq 0.07 $. Then, the $ (i+1) $-th matrix estimate $ \signal_{i+1} $ of Algorithm 1 can be decomposed into a superposition of low-rank and sparse components as $ \signal_{i+1} = \mathbf{L}_{i+1} + \mathbf{M}_{i+1} $, satisfying the recursions:  
\begin{align}
\vectornormbig{\mathbf{L}^{\ast} - \mathbf{L}_{i+1}}_F &\leq \rho_{1}^{\mathcal{L}} \vectornormbig{\mathbf{L}^{\ast} - \mathbf{L}_i}_F + \rho_{1}^{\mathcal{M}} \vectornormbig{\mathbf{M}^{\ast} - \mathbf{M}_i}_F + \gamma_1 \vectornormbig{\noise}_2 \nonumber \\
\vectornormbig{\mathbf{M}^{\ast} - \mathbf{M}_{i+1}}_F &\leq \rho_{2}^{\mathcal{L}} \vectornormbig{\mathbf{L}^{\ast} - \mathbf{L}_i}_F + \rho_{2}^{\mathcal{M}} \vectornormbig{\mathbf{M}^{\ast} - \mathbf{M}_i}_F + \gamma_2 \vectornormbig{\noise}_2 \nonumber  
\end{align} where $ \rho_1^{\mathcal{L}} = 0.1605 $, $ \rho_2^{\mathcal{L}} = 0.3431 $,  $ \rho_1^{\mathcal{M}} = 0.3376 $,  $ \rho_2^{\mathcal{M}} = 0.1414 $,  $ \gamma_1 = 4.36 $ and, $ \gamma_2 = 4.45$.  
\end{theorem} To compare with state-of-the-art approaches, \cite{sparcs} provides the following constants for the same RIP assumptions: $ \rho_1^{\mathcal{L}} = 0.479 $, $ \rho_2^{\mathcal{L}} = 0.474 $,  $ \rho_1^{\mathcal{M}} = 0.47 $,  $ \rho_2^{\mathcal{M}} = 0.324 $,  $ \gamma_1 = 6.68 $ and, $ \gamma_2 = 6.88$. 
We note here that the above theorem holds if and only if the intermediate estimates $\mathbf{L}_i$ and $\mathbf{M}_i$, $\forall i$, satisfy Lemma \ref{lemma:1}. Unfortunately, we cannot guarantee that $\mathbf{L}_i$ and $\mathbf{M}_i$ are uniformly bounded or have random support set patterns, respectively, at each iteration for arbitrary problem configurations. Although the potential optimization problem is non-convex, recent works on non-convex optimization \cite{attouch2010proximal, chouzenoux2011majorize} establish mild conditions on the objective function and the regularization terms, that are satisfied in our setting, under which a stationary point to a non-convex problem can be obtained using memory-less or memory-based projected gradient descent methods. 

Next, we sketch the proof of Theorem \ref{theorem:1} in a modular fashion and use key ingredients to analyze our \textsc{Matrix ALPS} algorithm.   

\subsection{Subspace and support exploration}  

\begin{lemma}[Active subspace expansion]\label{lemma:act_subspace_exp} At each iteration, the Active Subspace Expansion step (Step 4) captures information contained in the true matrix $ \mathbf{L}^{\ast} $ with $ \mathcal{L}^\ast \leftarrow \rm{ortho}(\mathbf{L}^{\ast}) $, such that:  
\begin{align}
\vectornormbig{\mathcal{P}_{\mathcal{L}^\ast \setminus \mathcal{S}_i^{\mathcal{L}}}&(\mathbf{L}^{\ast} - \mathbf{L}_i)}_F \leq (2\delta_{2\rank}(\sensing) + 2\delta_{3\rank}(\sensing))\vectornormbig{\mathbf{L}^{\ast} - \mathbf{L}_i}_F + 2\delta_{2\rank + 2\sparsity}(\sensing)\vectornormbig{\mathbf{M}^{\ast} - \mathbf{M}_i}_F + \sqrt{2(1+\delta_{2\rank}(\sensing))}\vectornormbig{\noise}_2. \nonumber  
\end{align} 
\end{lemma}

Lemma \ref{lemma:act_subspace_exp} states that, at each iteration, the Active subspace expansion step identifies a 2$ \rank $ rank subspace in $ \mathbb{R}^{\dimension} $ such that the amount of unrecovered energy of $ \mathbf{L}^\ast $---i.e., the projection of $ \mathbf{L}^{\ast}$ onto the orthogonal subspace of $ \rm{span}(\mathcal{S}_i^{\mathcal{L}}) $---is bounded as shown above. Similarly, the next Corollary holds for the sparse estimation part:

\begin{corollary}[Active support expansion]\label{lemma:act_support_exp} At each iteration, the Active Support Expansion step (Step 5) captures information contained in the true matrix $ \mathbf{M}^{\ast} $ with $ \mathcal{M}^\ast \leftarrow \rm{supp}(\mathbf{M}^{\ast}) $, such that:  
\begin{align}
\vectornormbig{\big( \mathbf{M}^{\ast} - \mathbf{M}_i\big)_{\mathcal{M}^{\ast} \setminus \mathcal{S}_i^{\mathcal{M}}} }_F
&\leq (\delta_{2\sparsity}(\sensing) + \delta_{4\sparsity}(\sensing))\vectornormbig{\mathbf{M}^{\ast} - \mathbf{M}_i}_F + \big(\delta_{2\rank + \sparsity}(\sensing) + \delta_{2\rank + 2\sparsity}(\sensing)\big)\vectornormbig{\mathbf{L}^{\ast} - \mathbf{L}_i}_F \nonumber \\ &+ \sqrt{2(1+\delta_{4\sparsity}(\sensing))}\vectornormbig{\noise}_2. \nonumber  
\end{align}
\end{corollary}  

\subsection{Least-squares estimates over low rank subspaces}   

\begin{lemma}[Least-squares error norm reduction over a low-rank subspace]\label{lemma:leastsquares_lowrank} Let $\mathcal{S}_{i}^{\mathcal{L}}$ be a set of orthonormal, rank-1 matrices such that $\rm{span}$ $(\mathcal{S}_i^{\mathcal{L}})$ $\leq 2 \rank $. Then, the rank-$2\rank$ solution $\mathbf{V}_i^{\mathcal{L}}$ in Step 7 identifies most of the energy of $\mathcal{L}^{\ast} $ over $\mathcal{S}_i^{\mathcal{L}}$ such that:  
\begin{align}
&\vectornormbig{\mathbf{V}_i^{\mathcal{L}} - \mathbf{L}^{\ast}}_F \nonumber \leq \frac{1}{\sqrt{1-\delta_{3\rank}^2(\sensing)}} \vectornormbig{\mathcal{P}_{\mathcal{S}_i^{\mathcal{L}}}^{\bot}(\mathbf{V}_i^{\mathcal{L}} - \mathbf{L}^{\ast})}_F + \frac{(1+2\delta_{2\rank}(\sensing))}{1 - \delta_{3\rank}^2(\sensing)}\Big(\delta_{2\rank + 2\sparsity}(\sensing)\vectornormbig{\mathbf{M}^{\ast} - \mathbf{M}_i}_F + \sqrt{1+\delta_{2\rank}(\sensing)} \vectornormbig{\noise}_2\Big). \nonumber  
\end{align} 
\end{lemma} Assuming $ \sensing $ is well-conditioned over low-rank subspaces, the main complexity of this operation is dominated by the solution of a symmetric linear system of equations.
Using Lemma \ref{lemma:leastsquares_lowrank} and the following inequality:  
\begin{align}
\vectornorm{\mathbf{L}_{i+1} - \mathbf{V}_i^{\mathcal{L}}}_F \leq \vectornormbig{\mathcal{P}_{\mathcal{S}_i^{\mathcal{L}}}(\mathbf{V}_i^{\mathcal{L}} - \mathbf{L}^{\ast})}_F \leq \vectornormbig{\mathbf{V}_i^{\mathcal{L}} - \mathbf{L}^{\ast}}_F, \nonumber \vspace{0.2cm}
\end{align} which is due to the best rank-$ \rank $ subspace selection on $\mathbf{V}_i^{\mathcal{L}}$ (Step 8), the following inequality holds true:  
\begin{align}
\vectornormbig{\mathbf{L}_{i+1} - \mathbf{L}^{\ast}}_F &\leq \sqrt{\frac{1+3\delta_{3\rank}^2(\sensing)}{1-\delta_{3\rank}^2(\sensing)}} \vectornormbig{\mathcal{P}_{\mathcal{S}_i^{\mathcal{L}}}^{\bot}(\mathbf{V}_{i}^{\mathcal{L}} -  \mathbf{L}^{\ast})}_F + \Big(\sqrt{1 + 3\delta_{3\rank}^2(\sensing)}\cdot\frac{1+2\delta_{2\rank}(\sensing)}{1-\delta_{3\rank}^2(\sensing)} + \sqrt{3}\Big)\big( \delta_{2\sparsity + 2\rank}(\sensing) \vectornormbig{\mathbf{M}^{\ast} - \mathbf{M}_i}_F \nonumber \\ &+ \sqrt{1 + \delta_{2\sparsity}(\sensing)}\vectornormbig{\noise}_2\big). \label{eq:inequality1}  
\end{align}
Combining Lemma \ref{lemma:act_subspace_exp} with the inequality (\ref{eq:inequality1}), we obtain the first inequality in Theorem 1.  

\subsection{Least-squares estimates over sparse support sets}  
Using similar techniques descibed above for the sparse matrix estimate, we derive the following result:
\begin{corollary}[Least-squares error norm reduction over sparse support sets]\label{lemma:leastsquares_sparse} Let $\mathcal{S}_{i}^{\mathcal{M}} \subseteq \lbrace (i,j): i \in \lbrace 1, \dots, m \rbrace, ~ j \in \lbrace 1, \dots n \rbrace \rbrace $ be a $2\sparsity$-sparse index set. Then, the $2\sparsity$-sparse matrix $\mathbf{V}_i^{\mathcal{M}}$ (Step 10) identifies energy of $\mathbf{M}^{\ast} $ over $\mathcal{S}_i^{\mathcal{M}}$ such that:  
\begin{align}
&\vectornormbig{\mathbf{V}_i^{\mathcal{M}} - \mathbf{M}^{\ast}}_F \leq \frac{1}{\sqrt{1 - \delta_{4\sparsity}^2(\sensing)}} \vectornormbig{\big(\mathbf{V}_{i}^{\mathcal{M}} -  \mathbf{M}^{\ast}\big)_{(\mathcal{S}_{i}^{\mathcal{M}})^c}}_F +  \frac{(1+2\delta_{2\sparsity}(\sensing))}{1-\delta_{4\sparsity}^2(\sensing)} \big( \delta_{3\sparsity + 2\rank}(\sensing) \vectornormbig{\mathbf{L}^{\ast} - \mathbf{L}_i}_F + \sqrt{1 + \delta_{3\sparsity}(\sensing)} \vectornormbig{\noise}_2 \big). \nonumber
\end{align}
\end{corollary} In sequence, we follow the same motions to obtain an inequality analogous to (\ref{eq:inequality1}) for the sparse matrix estimate part.  

\section{The \textsc{Matrix ALPS} Framework}{\label{sec:accelerated}}

\begin{algorithm}[t]
   \caption{\textsc{Matrix ALPS} Instance}\label{algo: class}
\begin{algorithmic}[1]
   \STATE {\bfseries Input:} $\obs$, $\sensing$, $ \sensing^{\ast} $,  Tolerance $ \eta $, MaxIterations, $\tau_i,~ \forall i$
   \STATE {\bfseries Initialize:} $ \lbrace \mathbf{Q}_0, \mathbf{M}_0, \mathbf{L}_0 \rbrace \leftarrow 0 $, $ \lbrace \mathcal{L}_0, \mathcal{M}_0 \rbrace \leftarrow \lbrace \emptyset \rbrace $, $ i \leftarrow 0 $
   \STATE {\bfseries repeat} 
   \STATE \hspace{0.02cm} \textbf{Low rank matrix estimation:}
   \STATE \hspace{0.25cm} $\mathcal{D}_i^{\mathcal{L}} \leftarrow \rm{ortho}$ $\big(\mathcal{P}_{\rank}(\nabla f(\mathbf{Q}_{i}))\big) $
   \STATE \hspace{0.25cm} $ \mathcal{S}_i^{\mathcal{L}} \leftarrow \mathcal{D}_i^{\mathcal{L}} \cup \mathcal{L}_i$
   \STATE \hspace{0.25cm} $ \mathbf{V}_{i}^{\mathcal{L}} \leftarrow \mathbf{Q}_i^\mathcal{L} - \frac{\mu_i^{\mathcal{L}}}{2} \mathcal{P}_{\mathcal{S}_i^{\mathcal{L}}} \nabla f(\mathbf{Q}_i) $
   \STATE \hspace{0.25cm} $ \mathbf{L}_{i+1} \leftarrow \mathcal{P}_{\rank}(\mathbf{V}_{i}^{\mathcal{L}}) ~~\text{with }~  \mathcal{L}_{i+1} \leftarrow \rm{ortho}$ $(\mathbf{L}_{i+1}) $ 
   \STATE \hspace{0.25cm} $ \mathbf{Q}_{i+1}^{\mathcal{L}} \leftarrow \mathbf{L}_{i+1} + \tau_i( \mathbf{L}_{i+1} - \mathbf{L}_i ) $
   \STATE \hspace{0.25cm} $ \mathbf{Q}_{i+1} \leftarrow \mathbf{Q}_{i+1}^{\mathcal{L}} + \mathbf{Q}_{i}^{\mathcal{M}} $

   \STATE \hspace{0.02cm} \textbf{Sparse matrix estimation:}
   \STATE \hspace{0.25cm} $\mathcal{D}_i^{\mathcal{M}} \leftarrow \rm{supp}$ $\big(\mathcal{P}_{\Sigma_{\sparsity}}(\nabla f(\mathbf{Q}_{i+1}))\big) $
   \STATE \hspace{0.25cm} $ \mathcal{S}_i^{\mathcal{M}} \leftarrow \mathcal{D}_i^{\mathcal{M}} \cup \mathcal{M}_i$
   \STATE \hspace{0.25cm} $ (\mathbf{V}_{i}^{\mathcal{M}})_{\mathcal{S}_i^{\mathcal{M}}} \leftarrow (\mathbf{Q}_i^\mathcal{M})_{\mathcal{S}_i^{\mathcal{M}}} - \frac{\mu_i^{\mathcal{M}}}{2} (\nabla f(\mathbf{Q}_{i+1}))_{\mathcal{S}_i^{\mathcal{M}}} $
   \STATE \hspace{0.25cm} $ \mathbf{M}_{i+1} \leftarrow \mathcal{P}_{\Sigma_{\sparsity}}(\mathbf{V}_{i}^{\mathcal{M}}) ~~\text{with }~  \mathcal{M}_{i+1} \leftarrow \rm{supp}$ $(\mathbf{M}_{i+1}) $ 
   \STATE \hspace{0.25cm} $ \mathbf{Q}_{i+1}^{\mathcal{M}} \leftarrow \mathbf{M}_{i+1} + \tau_i( \mathbf{M}_{i+1} - \mathbf{M}_i ) $
   
   \STATE \hspace{0.25cm} $\mathbf{Q}_{i+1} \leftarrow \mathbf{Q}_{i+1}^{\mathcal{L}} + \mathbf{Q}_{i+1}^{\mathcal{M}} $
   \STATE \hspace{0.02cm} $ i \leftarrow i + 1 $
   \STATE {\bfseries until} $\vectornorm{\signal_{i} - \signal_{i-1}}_2 \leq \eta \vectornorm{\signal_{i}}_2 $ or MaxIterations.
\end{algorithmic}
\end{algorithm}

To accelerate the convergence speed of SpaRCS, we propose \textsc{Matrix ALPS} algorithm based on acceleration techniques from convex analysis \cite{nesterov, KyrillidisCevherRecipes}. At each iteration, we leverage both low rank and sparse matrix estimates from previous iterations to form a gradient surrogate with low-computational cost. Then, we update the current estimates using memory to gain momentum in convergence as proposed in Nesterov's optimal gradient methods. A key ingredient is the selection of the momentum term $\tau$---constant and adaptive momentum selection strategies can be found in \cite{KyrillidisCevherRecipes}. We reserve the analysis for the adaptive case for an extended paper.

To further improve the convergence speed, we replace the least-squares optimization steps with first-order gradient descent updates---the step size $\mu_i^{\mathcal{L}},~\mu_i^{\mathcal{M}}$ selections follow from \cite{KyrillidisCevherRecipes}.

The best projection of an arbitrary matrix onto the set of low rank matrices requires sophisticated matrix decompositions such as Singular Value Decomposition (SVD). Using the Lanczos approach, we require $ O(\rank m n) $ arithmetic operations to compute a rank-$ \rank $ matrix approximation for a given constant accuracy---a prohibitive time-complexity that does not scale well for many practical applications. Alternatives to SVD can be found in \cite{halko2009finding, zhou2011godec}. Furthermore, \cite{KyrillidisCevherMatrixRecipes} includes $ \epsilon $-approximate low rank matrix projections in the recovery process and study their effects on the convergence.

The following theorem characterizes Algorithm 2 for the noiseless case using a constant momentum step size selection strategy.

\begin{theorem}\label{theorem:2} Let $\sensing: \mathbb{R}^{\dimension} \rightarrow \mathbb{R}^{\numsam} $ be a linear operator satisfying rank-RIP and sparse-RIP with constants $\delta_{4\rank}(\sensing) \leq 0.09$ and $\delta_{4\sparsity}(\sensing) \leq 0.095$, respectively. Furthermore, assume constant momentum step size selection with $\tau_i = 1/4, ~\forall i$. We consider the noiseless case where the set of observations satisfy $\obs = \sensing \bestsignal $ for $\bestsignal:= \mathbf{L}^{\ast} + \mathbf{M}^{\ast} $ as defined in \textsc{Problem}. Then, Algorithm 2 satisfies the following second-order linear system:
\begin{align} 
\mathbf{x}(i+1) \leq (1+\tau)\mathbf{\Delta} \mathbf{x}(i) + \tau \mathbf{\Delta} \mathbf{x}(i-1), \label{memory_original_inequality}
\end{align} where $\mathbf{x}(i) := \begin{bmatrix} \vectornormbig{\mathbf{L}_i - \mathbf{L}^{\ast}}_F \\ \vectornormbig{\mathbf{M}_i - \mathbf{M}^{\ast}}_F \end{bmatrix}$ and $\mathbf{\Delta} := \begin{bmatrix} \Delta_{11} & \Delta_{12} \\ \Delta_{21} & \Delta_{22} \end{bmatrix} $ depends on RIP constants $\delta_{4\rank}(\sensing) $ and $\delta_{4\sparsity}(\sensing)$. Furthermore, the above inequality can be transformed into the following first-order linear system:
\begin{align}
\mathbf{w}(i+1) \leq \underbrace{\begin{bmatrix} (1+\tau)\mathbf{\Delta} & \tau \mathbf{\Delta} \\ \mathbb{I} & \mathbf{0} \end{bmatrix}^i}_{\widehat{\mathbf{\Delta}}} \mathbf{w}(0),  \label{recursive}
\end{align} for $\mathbf{w}(i) := [ \mathbf{x}(i+1) ~~\mathbf{x}(i)]^T$. We observe that $\lim_{i \rightarrow \infty} \mathbf{w}(i) = \mathbf{0} $ since $|\lambda_j(\widehat{\mathbf{\Delta}})| \leq 1, ~\forall j $.
\end{theorem} Due to space constraints, we reserve the proof as well as the noisy analog of Theorem 2 for an extended version of the paper.

\section{Experiments}{\label{section:experiments}} 

\begin{figure*}[!]
\begin{center}
\begin{tabular}{|c|c|c|c|c|c|c|c|c|c|c|c|c|c|c|c|c|c}
\multicolumn{1}{c}{$m \times n$} & \multicolumn{1}{c}{$\rank$} & \multicolumn{1}{c|}{$\vectornormbig{\noise}_2$} & \multicolumn{5}{|c}{Iterations} & \multicolumn{5}{c}{Relative Error $:= \frac{\vectornormbig{\widehat{\signal} - \bestsignal}_F}{\vectornormbig{\bestsignal}_F} ~~(10^{-3})$} & \multicolumn{5}{c}{Time (sec)}\\
\hline
\multicolumn{1}{c}{$200 \times 400$} & \multicolumn{1}{c}{$5$} & \multicolumn{1}{c|}{0} & \multicolumn{5}{|c}{$29 ~/ ~24 / - / 46 / 11$} & \multicolumn{5}{c}{$0.134 / 0.18 / 0.002 / 0.78 / 0.04$} & \multicolumn{5}{c}{$ 2.26 / 0.27 / 0.95 / 0.36 / \mathbf{0.21} $} \\
\multicolumn{1}{c}{$200 \times 400$} & \multicolumn{1}{c}{$5$} & \multicolumn{1}{c|}{$10^{-2}$} & \multicolumn{5}{|c}{$29 / 24 / - / 45 / 11$} & \multicolumn{5}{c}{$0.127 / 0.164 / 0.01 / 0.76 / 0.05 $} & \multicolumn{5}{c}{$ 2.16 / 0.26 / 0.96 / 0.36 / \mathbf{0.23} $} \\
\multicolumn{1}{c}{$200 \times 400$} & \multicolumn{1}{c}{$10$} & \multicolumn{1}{c|}{$10^{-2}$} & \multicolumn{5}{|c}{$700 / 33 / - / 63 / 15$} & \multicolumn{5}{c}{$6.7 / 0.5 / 0.01 / 1.2 / 0.1 $} & \multicolumn{5}{c}{$ 36.38 / 0.45 / 1.13 / 0.64 / \mathbf{0.37} $} \\
\multicolumn{1}{c}{$200 \times 400$} & \multicolumn{1}{c}{$15$} & \multicolumn{1}{c|}{0} & \multicolumn{5}{|c}{$700 / 48 / - / 88 / 22$} & \multicolumn{5}{c}{$150 / 0.93 / 340/ 2.1 / 0.15 $} & \multicolumn{5}{c}{$ 98.12 / 0.82 / 1.29 / 1.08 / \mathbf{0.68} $} \\
\hline
\multicolumn{1}{c}{$1000 \times 5000$} & \multicolumn{1}{c}{$10$} & \multicolumn{1}{c|}{0} & \multicolumn{5}{|c}{$- / 22 / - / 30 / 6$} & \multicolumn{5}{c}{$- / 0.09 / 0.008 / 0.34 / 0.03$} & \multicolumn{5}{c}{$ - / 10.8 / 27.6 / 10.2 / \mathbf{5.5} $} \\
\multicolumn{1}{c}{$1000 \times 5000$} & \multicolumn{1}{c}{$50$} & \multicolumn{1}{c|}{$10^{-4}$} & \multicolumn{5}{|c}{$- / 24 / - / 48 / 10$} & \multicolumn{5}{c}{$- / 0.2 / 0.002 / 0.73 / 0.11$} & \multicolumn{5}{c}{$ - / 23.4 / 171.37 / 35.5 / \mathbf{17.2} $} \\
\multicolumn{1}{c}{$1000 \times 5000$} & \multicolumn{1}{c}{$120$} & \multicolumn{1}{c|}{0} & \multicolumn{5}{|c}{$- / 63 / - / 118 / 26$} & \multicolumn{5}{c}{$- / 0.52 / 0.07 / 1.22 / 0.077$} & \multicolumn{5}{c}{$ - / 139 / 501 / 228 / \mathbf{101} $} \\
\hline
\end{tabular}
\end{center}
\caption{Comparison table for the matrix completion problem. Table depicts median values over 50 Monte-Carlo iterations. To separate the results, we use ``$\big/$''. The list of algorithms includes: SpaRCS \cite{sparcs} $\big/$ ALM \cite{ALM} $\big/$ GROUSE \cite{GROUSE} $\big/$ SVP \cite{SVP} $\big/$ \textsc{Matrix ALPS}. Bold numbers highlight the fastest convergenvce in execution time. ``$-$`` denotes either no information or not applicable due to slow convergence.}
\end{figure*}

\textbf{Robust matrix completion:\footnote{Codes are available for MATLAB at {\it http://lions.epfl.ch/MatrixALPS}}} The rank-$\rank$ $\bestsignal \in \mathbb{R}^{\dimension}$ is synthesized as $\bestsignal:=\mathbf{U}\mathbf{R}^T$ where $\mathbf{U} \in \mathbb{R}^{m \times \rank} $ and $\mathbf{R} \in \mathbb{R}^{n \times \rank} $ and $\vectornormbig{\bestsignal}_F = 1$. We subsample $\bestsignal$ by observing $\numsam = 0.3 mn $ entries, drawn uniformly at random. The set of observations satisfies: $ \obs = \boldsymbol{\mathcal{A}}_{\Omega}\bestsignal + \noise $. Here, $\Omega$ denotes the set of ordered pairs that represent the coordinates of the observable entries and $\boldsymbol{\mathcal{A}}_{\Omega}$ denotes the linear operator (mask) that subsamples a matrix according to $\Omega$. 

We generate various problem configurations, both for noisy and noiseless settings. All the algorithms are tested for the same signal-matrix-noise realizations and use the same tolerance parameter $\eta = 10^{-4}$. For fairness, we modified all the algorithms so that {\it they exploit the true rank}. For low-rank projections, we use $\rm{PROPACK} $ package \cite{propack}, except \cite{GROUSE} which is SVD-less. We changed the maximum number of cycles in \cite{GROUSE} from $150 $ to $30$ to improve its speed. A summary of the results can be found in Fig. 1. We observe that \textsc{Matrix ALPS} has better phase transition performance. A complete comparison using randomized, low-rank projection schemes in \textsc{Matrix ALPS} is provided in the extended paper. 

\textbf{RPCA:} We consider the problem of background subtraction in video sequences: static brackground scenes are considered low-rank while moving foreground objects are sparse data. Using the complete set of measurements, this problem falls under the RPCA framework. We apply the GoDec algorithm \cite{zhou2011godec} and the \textsc{Matrix ALPS} scheme on a 144 x 176 x 200 video sequence. Both solvers use the same low-rank projection operators based on randomized QR factorization ideas \cite{halko2009finding, zhou2011godec}. Representative results are depicted in Fig. 2.

\begin{figure}[t]
\centering
\begin{minipage}{0.3\linewidth}
\centering {Original}
\end{minipage}
\begin{minipage}{0.3\linewidth}
\centering {Low rank}
\end{minipage}
\begin{minipage}{0.3\linewidth}
\centering {Sparse}
\end{minipage}

\includegraphics[width = 0.27\linewidth]{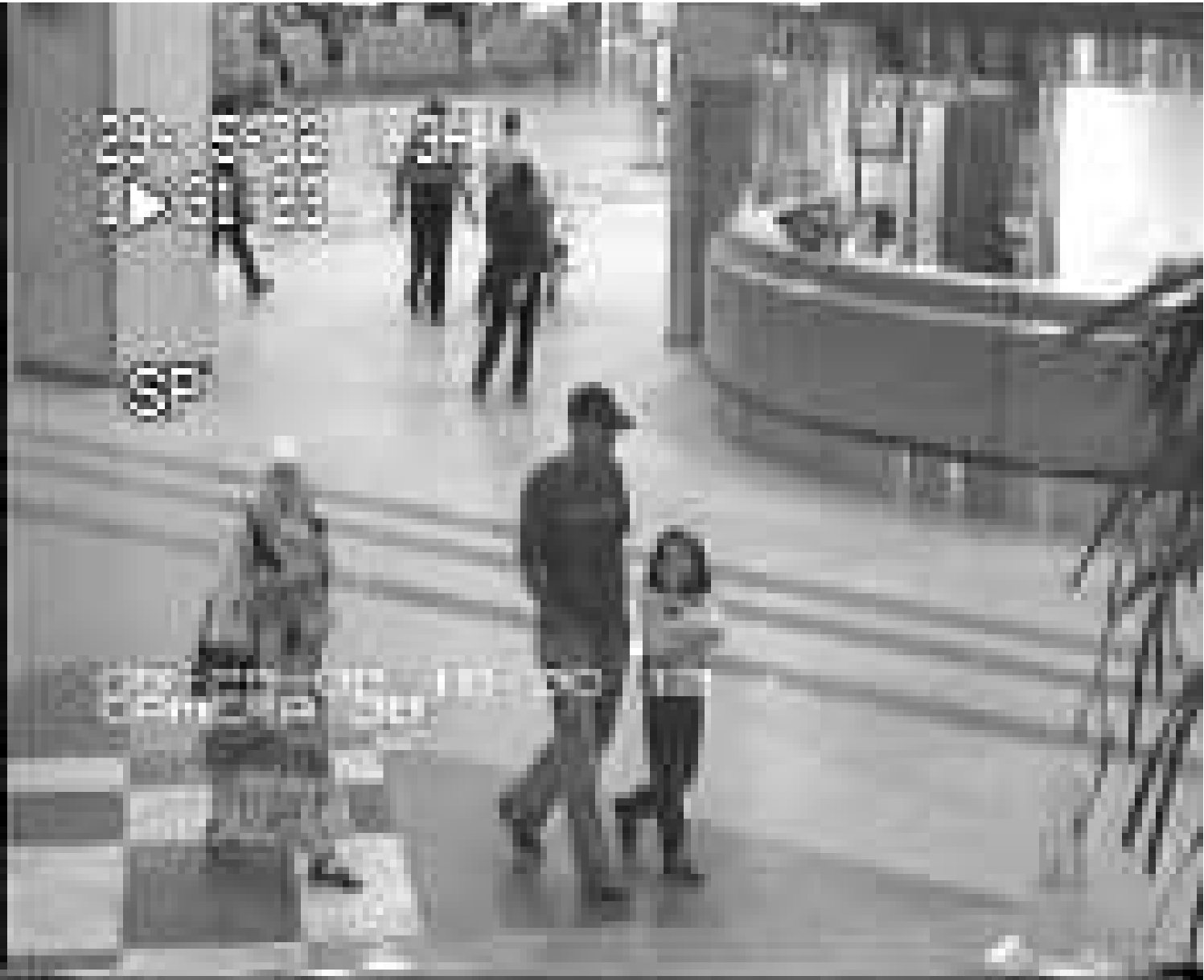} \hspace{0.2cm}
\includegraphics[width = 0.27\linewidth]{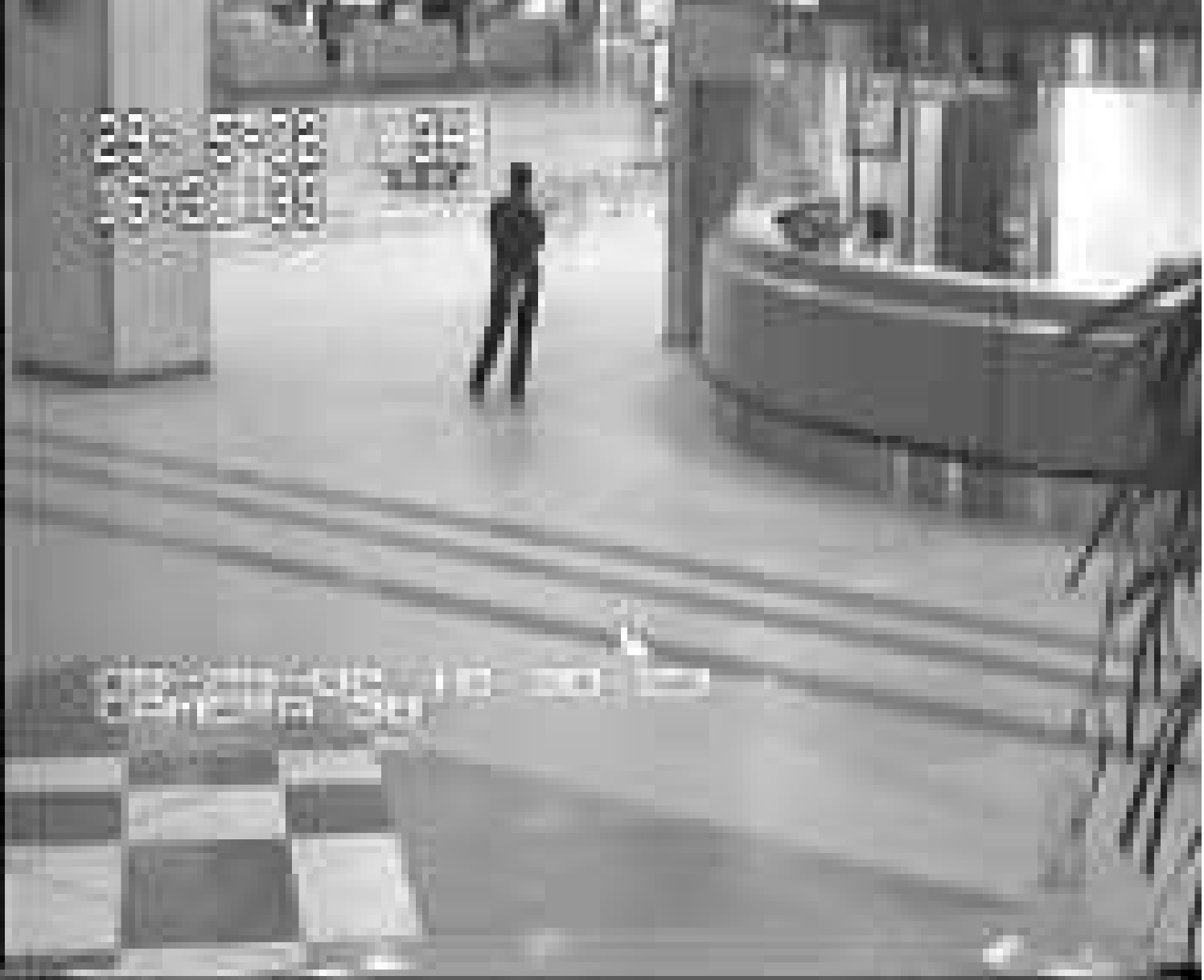} \hspace{0.2cm}
\includegraphics[width = 0.27\linewidth]{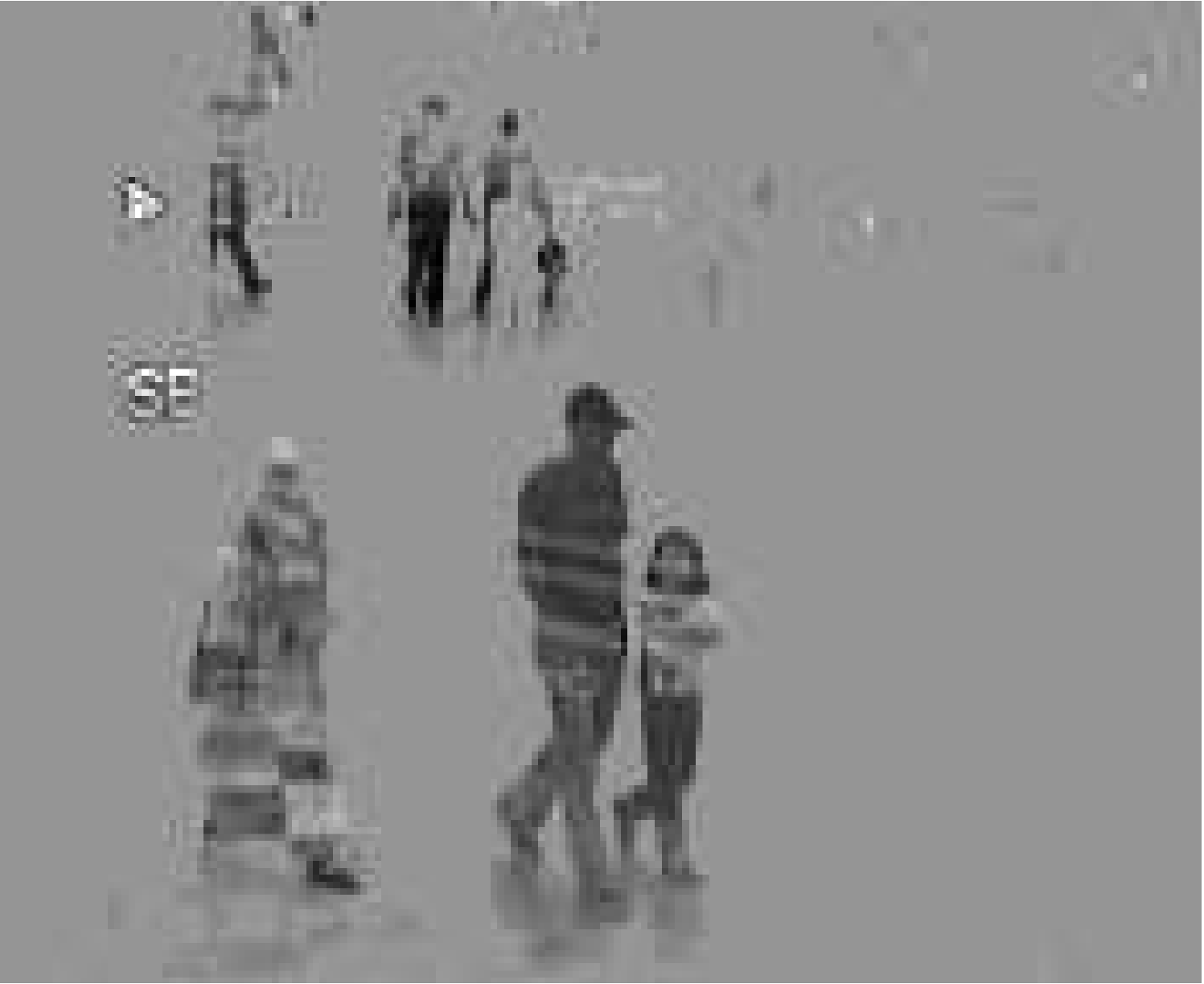} \\

\hspace{0.1cm}\includegraphics[width = 0.27\linewidth]{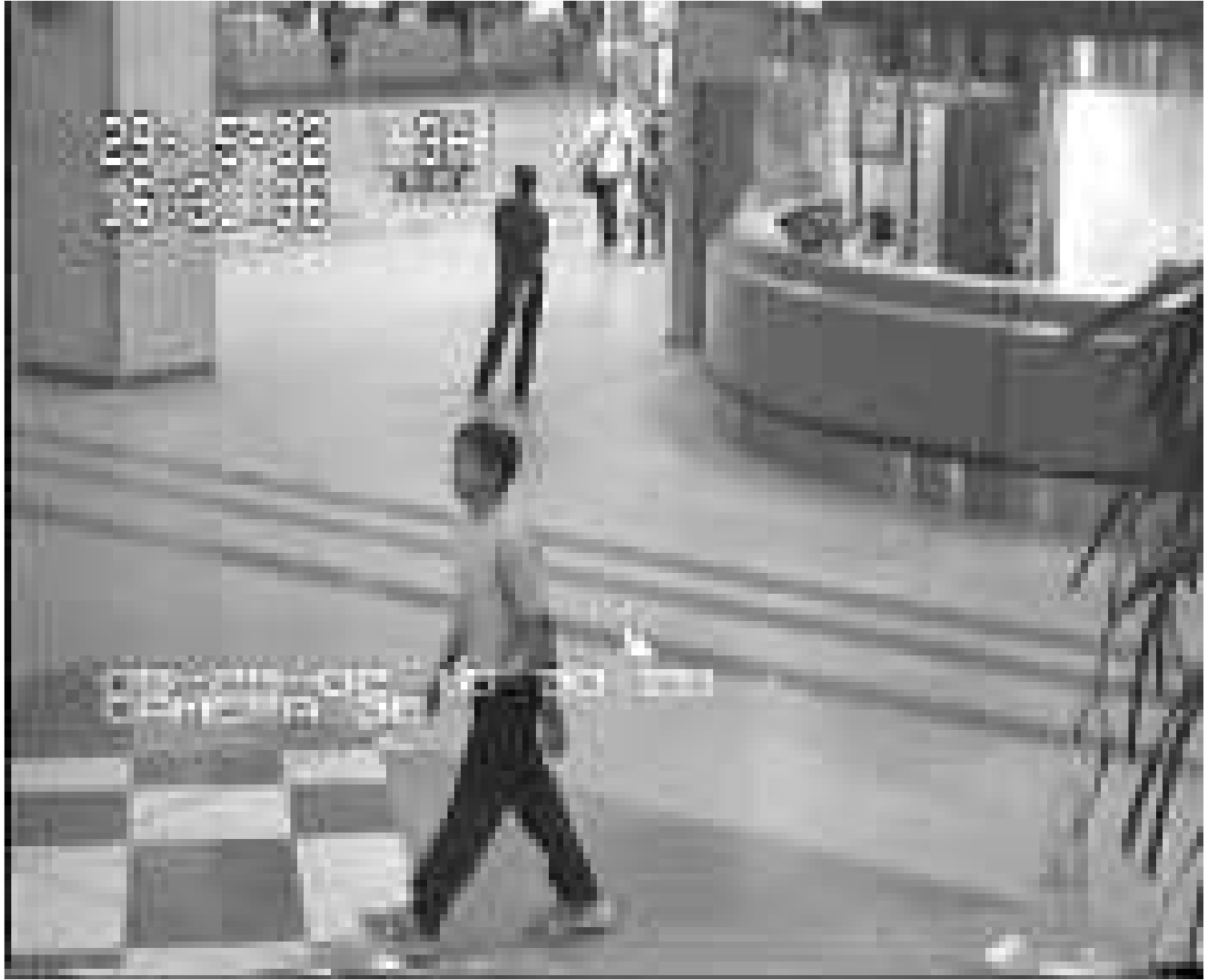}  \hspace{0.2cm}
\includegraphics[width = 0.27\linewidth]{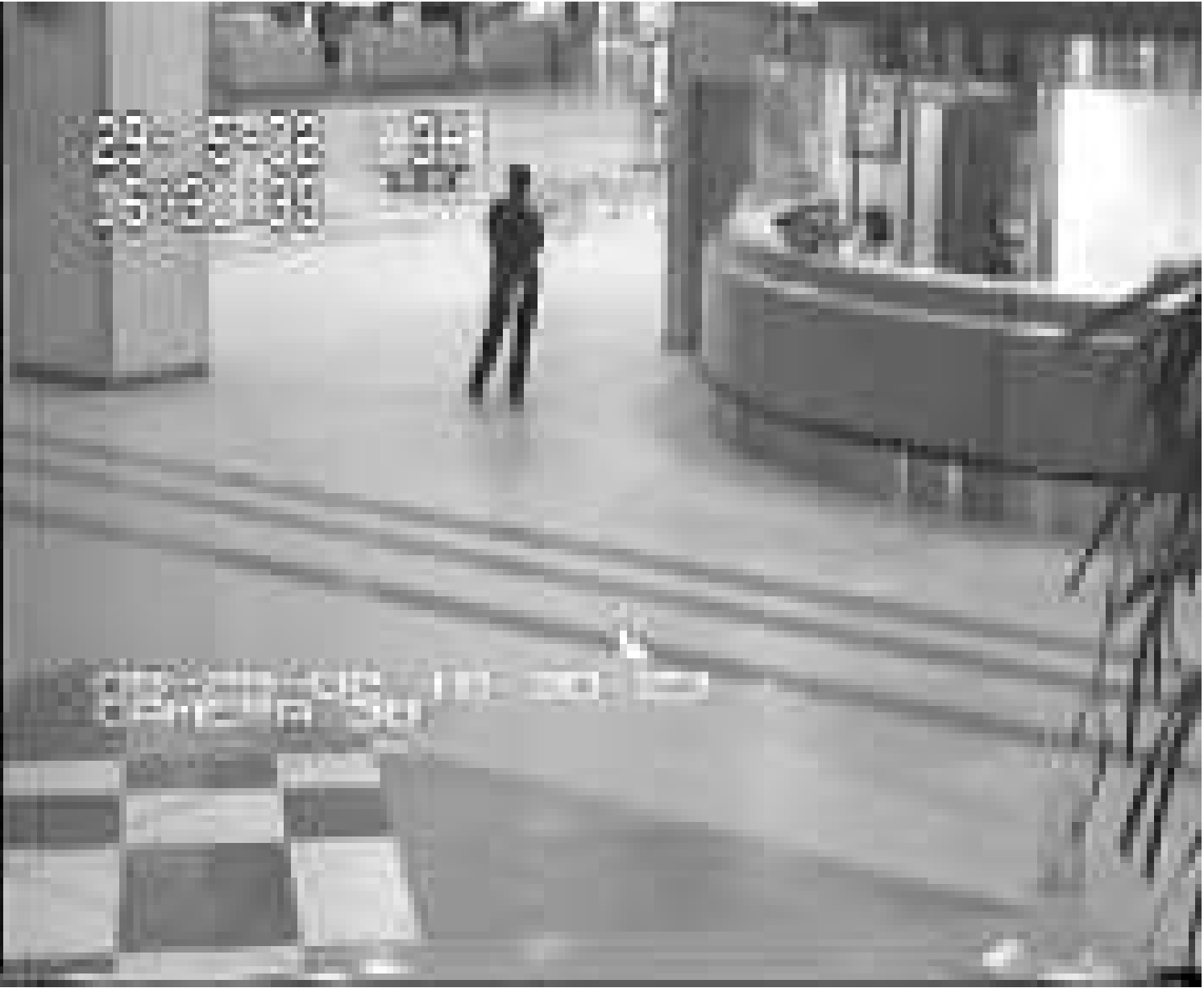} \hspace{0.2cm}
\includegraphics[width = 0.27\linewidth]{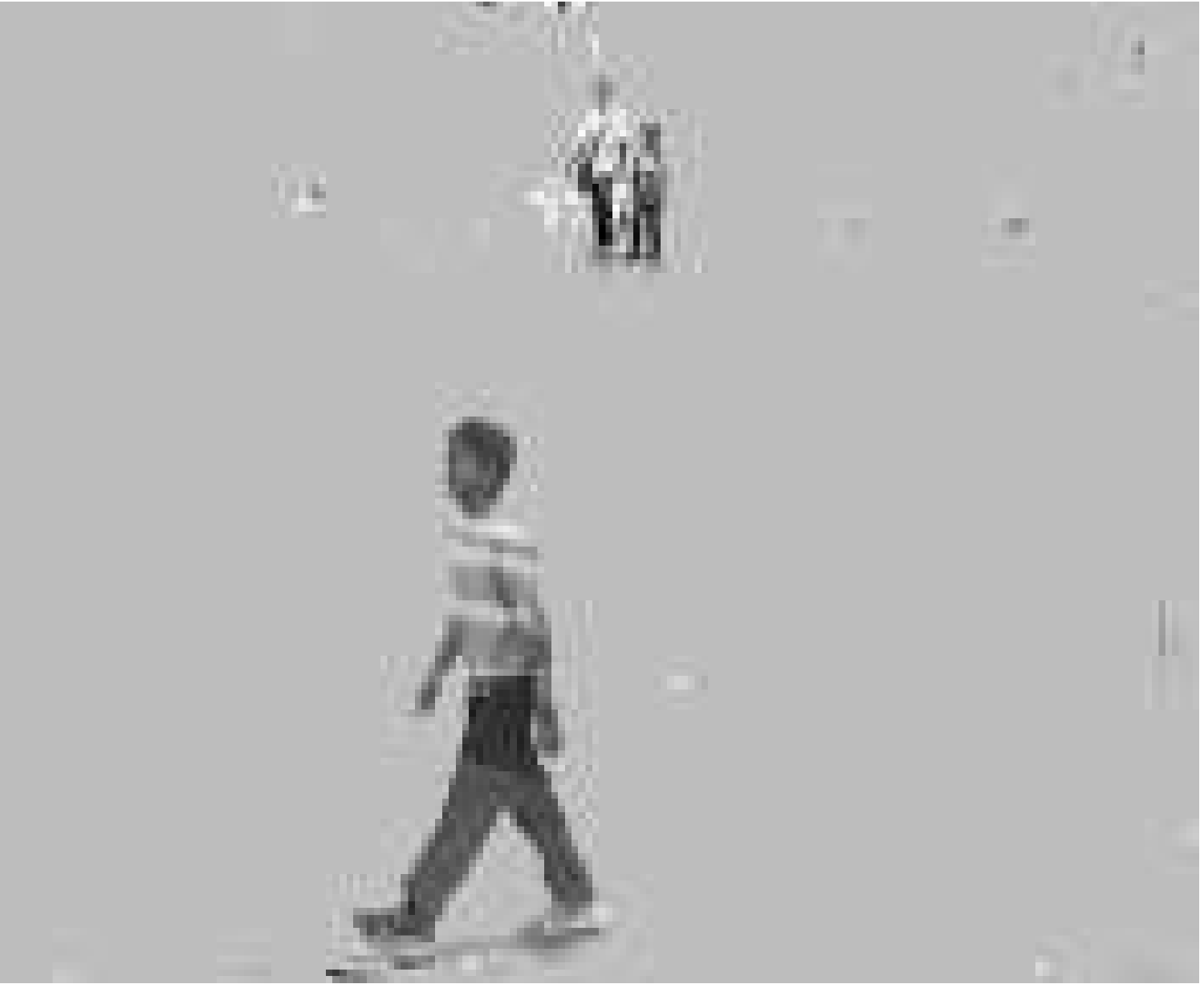} \vspace{-0.1cm}\\
GoDec \vspace{0.1cm} \\

\includegraphics[width = 0.27\linewidth]{original_1} \hspace{0.2cm}
\includegraphics[width = 0.27\linewidth]{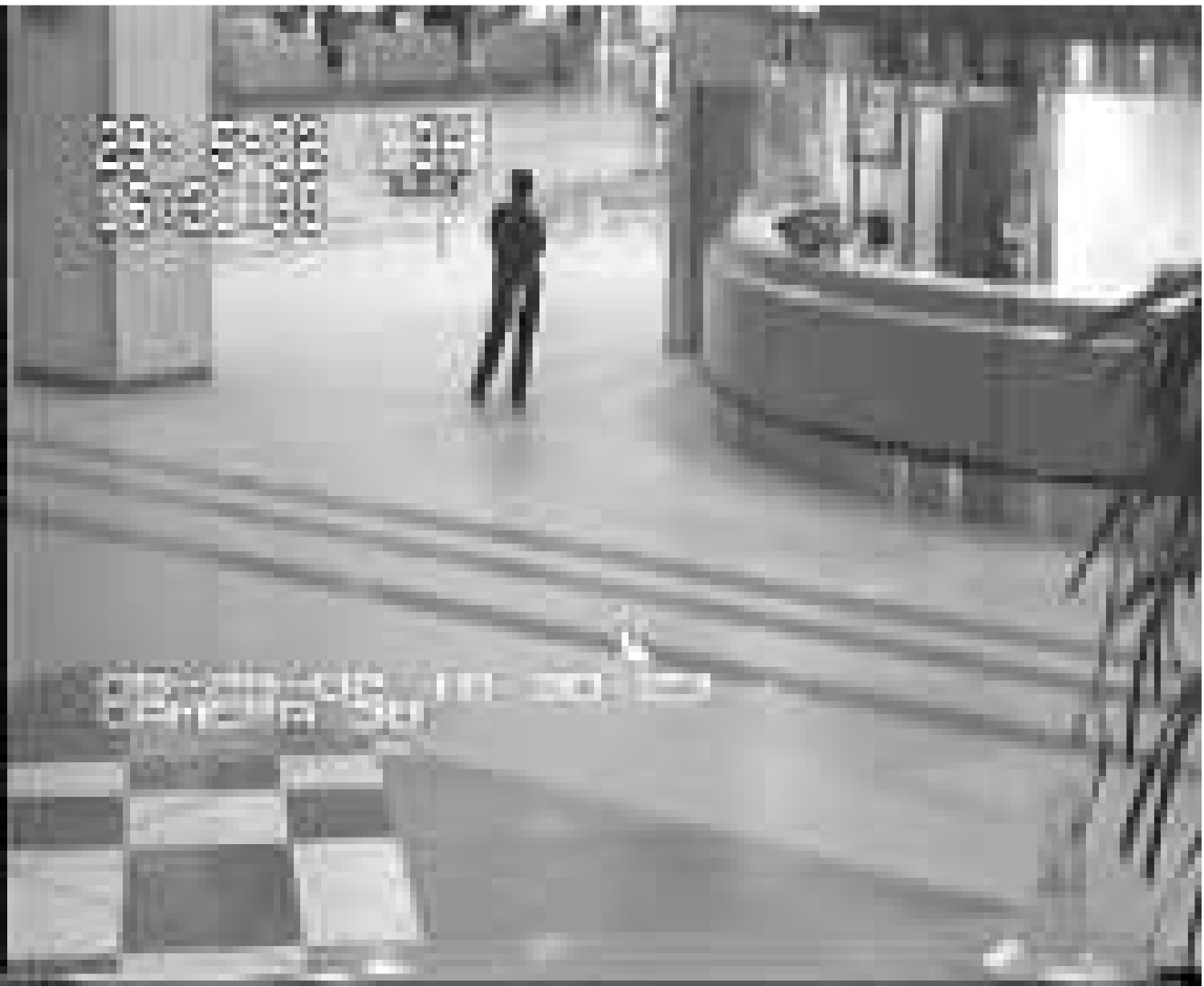} \hspace{0.2cm}
\includegraphics[width = 0.27\linewidth]{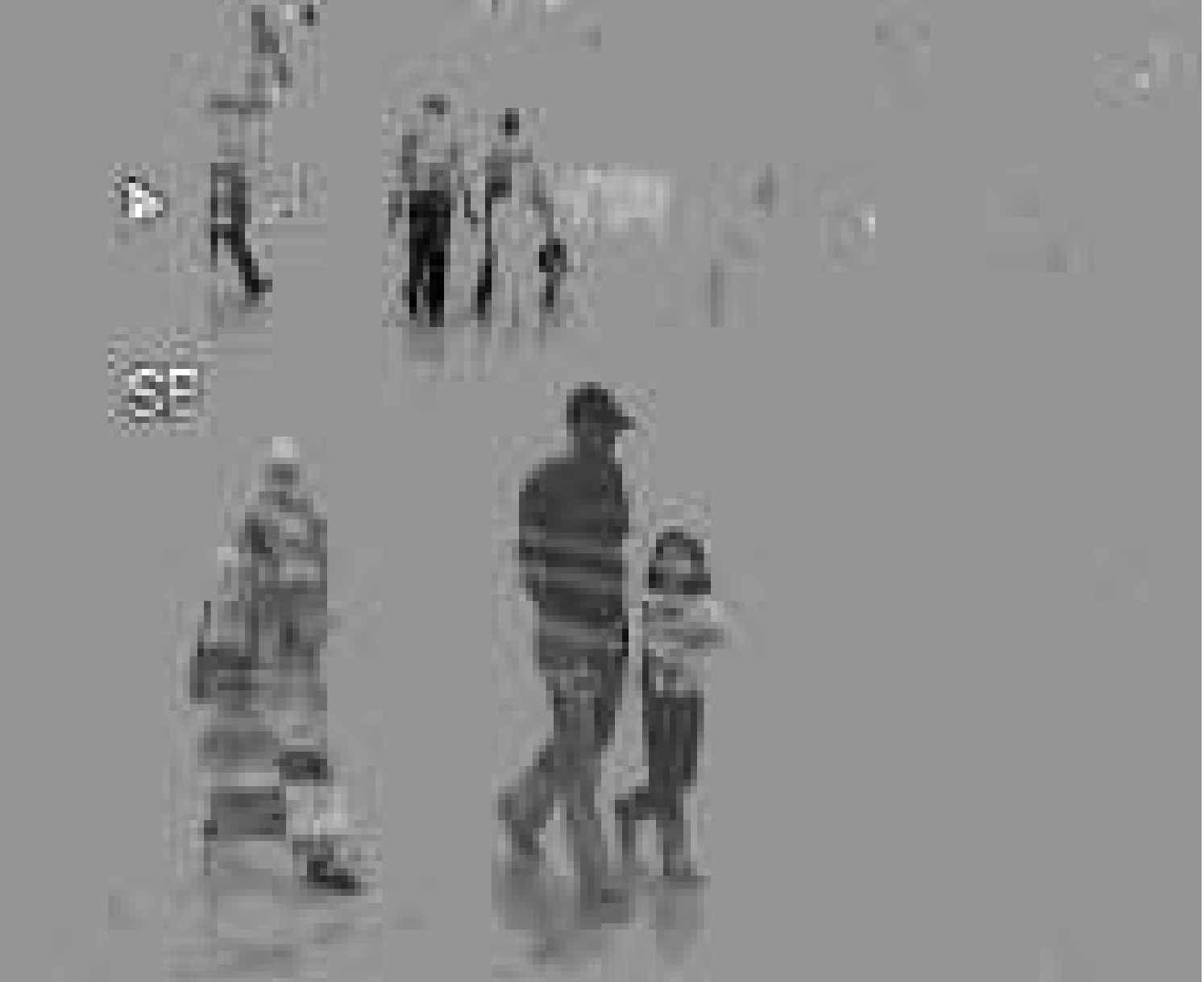} \\

\hspace{0.1cm}\includegraphics[width = 0.27\linewidth]{original_130} \hspace{0.2cm}
\includegraphics[width = 0.27\linewidth]{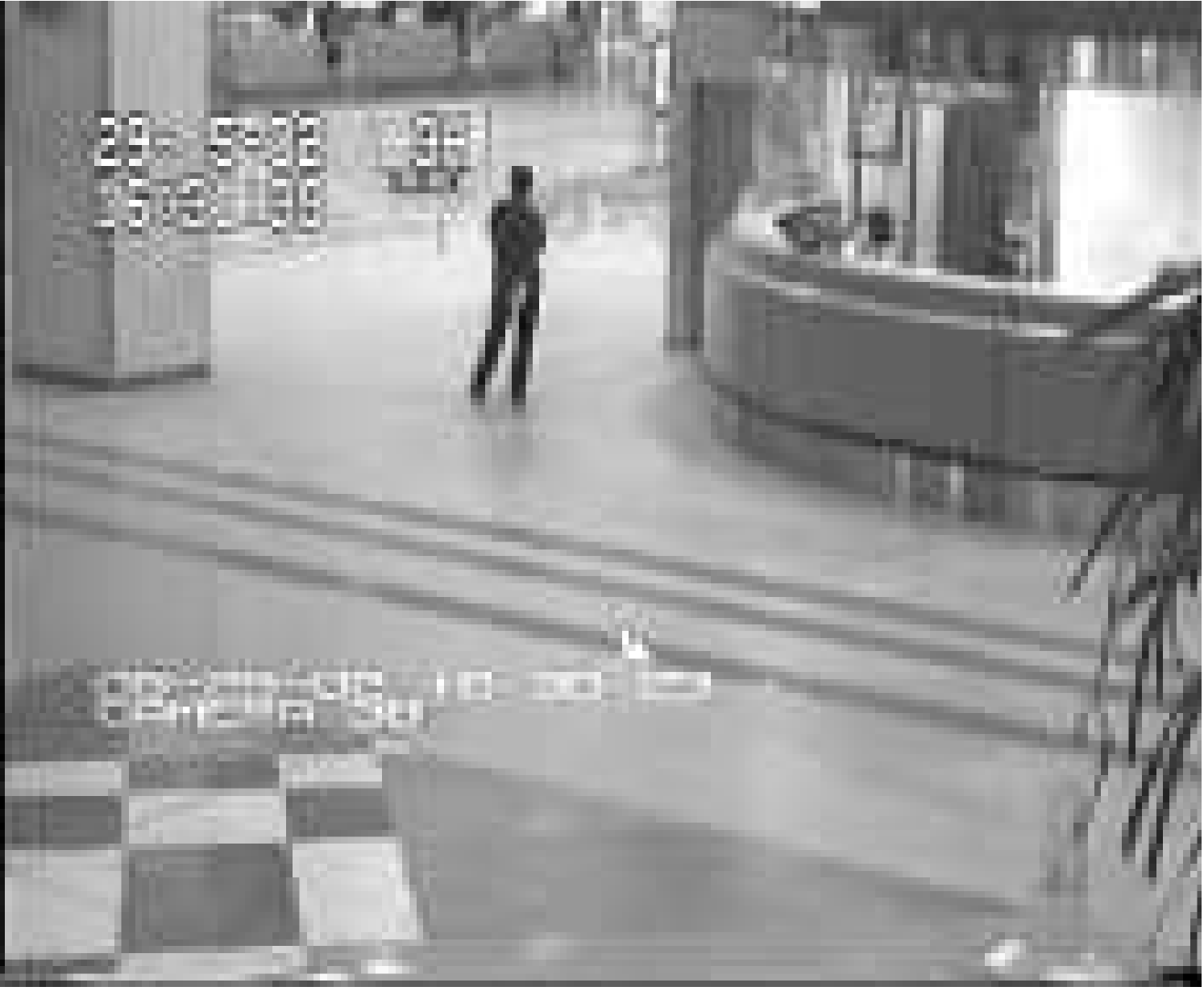} \hspace{0.2cm}
\includegraphics[width = 0.27\linewidth]{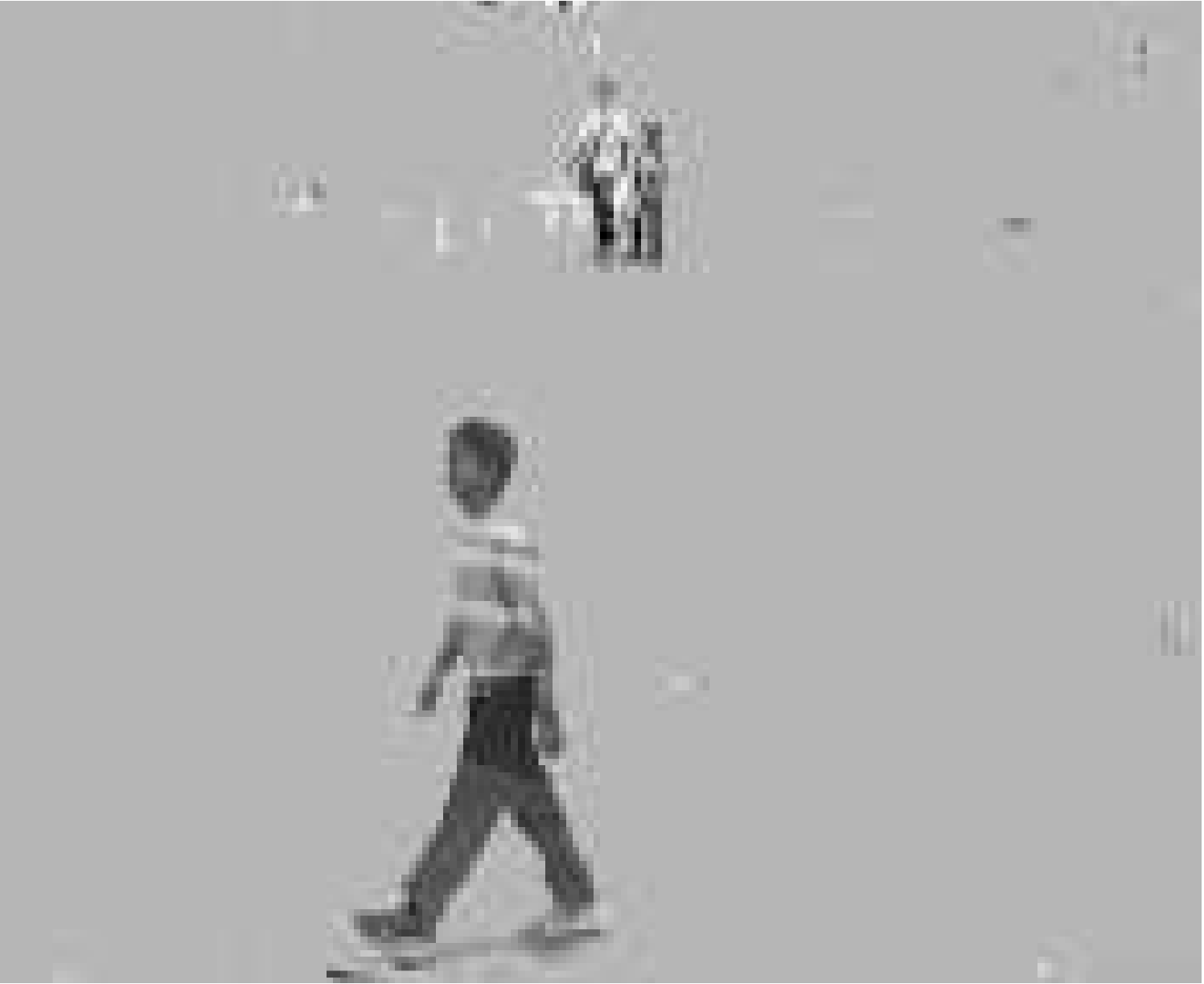} \vspace{-0.1cm} \\

\textsc{Matrix ALPS}

\caption{\small{Background subtraction in video sequence. Median execution times over 10 Monte-Carlo iterations. GoDec: $34.8$ sec---\textsc{Matrix ALPS}: $15.8$ sec.}} {\label{fig:real}}
\end{figure}
\section{conclusions}
We study the general problem of sparse plus low rank matrix recovery from incomplete and noisy data. In essence, the problem under consideration includes various low-dimensional models as special cases such as sparse signal reconstruction, affine rank minimization and robust PCA. Based on this algorithm, we derive improved conditions on the restricted isometry constants that guarantee the success of reconstruction. Furthermore, we show that the memory-based scheme provides great computational advantage over both the convex and the non-convex approaches. 

\section{Appendix}
\subsection{Proof of Lemma \ref{lemma:act_subspace_exp}}

Given $ \mathcal{L}^\ast:= \text{ortho}(\mathbf{L}^{\ast}) $, we define the following quantities: $ \mathcal{S}_i^{\mathcal{L}} := \mathcal{L}_i \cup \mathcal{D}_i^{\mathcal{L}},~ \widehat{\mathcal{S}}^{\mathcal{L}}_i := \mathcal{L}_i \cup \mathcal{L}^\ast $. Then:
\begin{align}
\mathcal{P}_{\mathcal{S}_i^{\mathcal{L}} \setminus \widehat{\mathcal{S}}_i^{\mathcal{L}}} = \mathcal{P}_{\mathcal{D}_i^{\mathcal{L}} \setminus (\mathcal{L}^\ast \cup \mathcal{L}_i)}, ~~\text{and}~~ \mathcal{P}_{\widehat{\mathcal{S}}^{\mathcal{L}}_i \setminus \mathcal{S}_i^{\mathcal{L}}} = \mathcal{P}_{\mathcal{L}^\ast \setminus (\mathcal{D}_i^{\mathcal{L}} \cup \mathcal{L}_i)} \label{ser:eq:00}.
\end{align} Since the subspace defined in $ \mathcal{D}_i^{\mathcal{L}} $ is the best rank-$ \rank $ subspace, orthogonal to the subspace spanned by $ \mathcal{L}_i $, the following holds true:
\begin{align}
\vectornormbig{\mathcal{P}_{\mathcal{D}_i^{\mathcal{L}} \setminus \mathcal{L}_i} \nabla f(\signal_{i})}_F^2 \geq \vectornormbig{\mathcal{P}_{\mathcal{L}^\ast \setminus \mathcal{L}_i} \nabla f(\signal_{i})}_F^2 \Rightarrow 
\vectornormbig{\mathcal{P}_{\mathcal{S}_i^{\mathcal{L}}} \nabla f(\signal_{i})}_F^2 \geq \vectornormbig{\mathcal{P}_{\widehat{\mathcal{S}}_i^{\mathcal{L}}} \nabla f(\signal_{i})}_F^2
\end{align} Removing the common subspaces in $ \mathcal{S}_i^{\mathcal{L}} $ and $ \widehat{\mathcal{S}}_i^{\mathcal{L}} $, we get
\begin{align}
\vectornormbig{\mathcal{P}_{\mathcal{S}_i^{\mathcal{L}} \setminus \widehat{\mathcal{S}}_i^{\mathcal{L}}} &\sensing^\ast \sensing (\mathbf{L}^{\ast} - \mathbf{L}_{i}) + \mathcal{P}_{\mathcal{S}_i^{\mathcal{L}} \setminus \widehat{\mathcal{S}}_i^{\mathcal{L}}} \sensing^\ast \sensing (\mathbf{M}^{\ast}- \mathbf{M}_{i}) + \mathcal{P}_{\mathcal{S}_i^{\mathcal{L}} \setminus \widehat{\mathcal{S}}_i^{\mathcal{L}}} \sensing^\ast \noise}_F \geq \nonumber \\ &\vectornormbig{\mathcal{P}_{\widehat{\mathcal{S}}_i^{\mathcal{L}} \setminus \mathcal{S}_i^{\mathcal{L}}} \sensing^\ast \sensing (\mathbf{L}^{\ast}- \mathbf{L}_{i}) + \mathcal{P}_{\widehat{\mathcal{S}}_i^{\mathcal{L}} \setminus \mathcal{S}_i^{\mathcal{L}}} \sensing^\ast \sensing (\mathbf{M}^{\ast} - \mathbf{M}_{i}) + \mathcal{P}_{\widehat{\mathcal{S}}_i^{\mathcal{L}} \setminus \mathcal{S}_i^{\mathcal{L}}} \sensing^\ast \noise}_F \label{ser:eq:01}
\end{align} On the left hand side, we have:
\begin{align}
\vectornormbig{\mathcal{P}_{\mathcal{S}_i^{\mathcal{L}} \setminus \widehat{\mathcal{S}}_i^{\mathcal{L}}} &\sensing^\ast \sensing (\mathbf{L}^{\ast} - \mathbf{L}_i) + \mathcal{P}_{\mathcal{S}_i^{\mathcal{L}} \setminus \widehat{\mathcal{S}}_i^{\mathcal{L}}} \sensing^\ast \sensing (\mathbf{M}^{\ast}- \mathbf{M}_{i}) + \mathcal{P}_{\mathcal{S}_i^{\mathcal{L}} \setminus \widehat{\mathcal{S}}_i^{\mathcal{L}}} \sensing^\ast \noise}_F \\ 
&\stackrel{(i)}{\leq} \vectornormbig{\mathcal{P}_{\mathcal{S}_i^{\mathcal{L}} \setminus \widehat{\mathcal{S}}_i^{\mathcal{L}}} \sensing^\ast \sensing (\mathbf{L}^{\ast} - \mathbf{L}_i)}_F +  \vectornormbig{\mathcal{P}_{\mathcal{S}_i^{\mathcal{L}} \setminus \widehat{\mathcal{S}}_i^{\mathcal{L}}} \sensing^\ast \noise}_F  + \vectornormbig{\mathcal{P}_{\mathcal{S}_i^{\mathcal{L}} \setminus \widehat{\mathcal{S}}_i^{\mathcal{L}}} \sensing^\ast \sensing (\mathbf{M}^{\ast} - \mathbf{M}_{i})}_F \label{ser:eq:02}
\end{align} where $ (i) $ due to triangle inequality over Frobenius metric norm. The first two terms in the above expression can be bounded using tools in \cite{KyrillidisCevherMatrixRecipes}. For the third term, we use Lemma 3.2 in \cite{sparcs} where we conclude that $\vectornormbig{\mathcal{P}_{\mathcal{S}_i^{\mathcal{L}} \setminus \widehat{\mathcal{S}}_i^{\mathcal{L}}} \sensing^\ast \sensing (\mathbf{M}^{\ast}- \mathbf{M}_{i})}_F \leq \delta_{2\rank + 2\sparsity}(\sensing) \vectornormbig{\mathbf{M}^{\ast} - \mathbf{M}_i}_F$. Thus:
\begin{align}
\vectornormbig{\mathcal{P}_{\mathcal{S}_i^{\mathcal{L}} \setminus \widehat{\mathcal{S}}_i^{\mathcal{L}}} &\sensing^\ast \sensing (\mathbf{L}^{\ast} - \mathbf{L}_i) + \mathcal{P}_{\mathcal{S}_i^{\mathcal{L}} \setminus \widehat{\mathcal{S}}_i^{\mathcal{L}}} \sensing^\ast \sensing (\mathbf{M}^{\ast}- \mathbf{M}_{i}) + \mathcal{P}_{\mathcal{S}_i^{\mathcal{L}} \setminus \widehat{\mathcal{S}}_i^{\mathcal{L}}} \sensing^\ast \noise}_F \\ &\leq 2\delta_{3\rank}(\sensing)\vectornormbig{\mathbf{L}^{\ast} - \mathbf{L}_i}_F + \delta_{2\rank + 2\sparsity}(\sensing)\vectornormbig{\mathbf{M}^{\ast} - \mathbf{M}_i}_F + \vectornormbig{\mathcal{P}_{\mathcal{S}_i^{\mathcal{L}} \setminus \widehat{\mathcal{S}}_i^{\mathcal{L}}} \sensing^\ast \noise}_F
\end{align}

Similarly, using ideas from \cite{KyrillidisCevherMatrixRecipes} for the right hand side,  we calculate:
\begin{align}
\vectornormbig{\mathcal{P}_{\mathcal{S}_i^{\mathcal{L}} \setminus \widehat{\mathcal{S}}_i^{\mathcal{L}}} &\sensing^\ast \sensing (\mathbf{L}^{\ast} - \mathbf{L}_{i}) + \mathcal{P}_{\mathcal{S}_i^{\mathcal{L}} \setminus \widehat{\mathcal{S}}_i^{\mathcal{L}}} \sensing^\ast \sensing (\mathbf{M}^{\ast}- \mathbf{M}_{i}) + \mathcal{P}_{\mathcal{S}_i^{\mathcal{L}} \setminus \widehat{\mathcal{S}}_i^{\mathcal{L}}} \sensing^\ast \noise}_F \\
&\geq \vectornormbig{\mathcal{P}_{\widehat{\mathcal{S}}_i^{\mathcal{L}} \setminus \mathcal{S}_i^{\mathcal{L}}} (\mathbf{L}^{\ast} - \mathbf{L}_{i})}_F - 2\delta_{2\rank}(\sensing)\vectornormbig{\mathbf{L}^{\ast} - \mathbf{L}_i}_F - \delta_{2\rank + 2\sparsity}(\sensing) \vectornormbig{\mathbf{M}^{\ast} - \mathbf{M}_i}_F - \vectornormbig{\mathcal{P}_{\widehat{\mathcal{S}}_i^{\mathcal{L}} \setminus \mathcal{S}_i^{\mathcal{L}}}\sensing^\ast \noise}_F \label{ser:eq:03}
\end{align} Combining the above inequalities, we get:
\begin{align}
&\vectornormbig{\mathcal{P}_{\mathcal{L}^\ast \setminus (\mathcal{D}_i^{\mathcal{L}} \cup \mathcal{L}_i)}(\mathbf{L}^{\ast} - \mathbf{L}_i)}_F \nonumber \\ 
&\leq (2\delta_{2\rank}(\sensing) + 2\delta_{3\rank}(\sensing))\vectornormbig{\mathbf{L}^{\ast} - \mathbf{L}_i}_F + 2\delta_{2\rank + 2\sparsity}(\sensing)\vectornormbig{\mathbf{M}^{\ast} - \mathbf{M}_i}_F + \vectornormbig{ \mathcal{P}_{(\widehat{\mathcal{S}}_i^{\mathcal{L}} \setminus \mathcal{S}_i^{\mathcal{L}}) \cup (\mathcal{S}_i^{\mathcal{L}} \setminus \widehat{\mathcal{S}}_i^{\mathcal{L}})} \sensing^\ast \noise}_F \\
 &\leq (2\delta_{2\rank}(\sensing) + 2\delta_{3\rank}(\sensing))\vectornormbig{\mathbf{L}^{\ast} - \mathbf{L}_i}_F + 2\delta_{2\rank + 2\sparsity}(\sensing)\vectornormbig{\mathbf{M}^{\ast} - \mathbf{M}_i}_F + \sqrt{2(1+\delta_{2\rank}(\sensing))}\vectornormbig{\noise}_2.
\end{align} To prove Corollary \ref{lemma:act_support_exp}, we follow the same ideas based on \cite{cosamp, foucart2010sparse, KyrillidisCevherMatrixRecipes}.

\subsection{Proof of Lemma \ref{lemma:leastsquares_lowrank}}

We observe that $ \vectornormbig{\mathbf{V}_i^{\mathcal{L}} - \mathbf{L}^{\ast}}_F^2 $ is decomposed as follows:
\begin{align}
\vectornormbig{\mathbf{V}_i^{\mathcal{L}} - \mathbf{L}^{\ast}}_F^2 = \vectornormbig{\mathcal{P}_{\mathcal{S}_i^{\mathcal{L}}} (\mathbf{V}_i^{\mathcal{L}} - \mathbf{L}^{\ast})}_F^2 + \vectornormbig{\mathcal{P}_{\mathcal{S}_i}^{\bot}(\mathbf{V}_i^{\mathcal{L}} - \mathbf{L}^{\ast})}_F^2. \label{eq:mALPS5:02}
\end{align}
$ \mathbf{V}_i^{\mathcal{L}} $ is the minimizer over the low-rank subspace spanned by $ \mathcal{S}_i^{\mathcal{L}} $ with $ \text{rank}(\text{span}(\mathcal{S}_i^{\mathcal{L}})) \leq 2\rank $. Using the optimality condition over the convex set $ \Theta = \lbrace \signal: \text{span}(\signal) \in \mathcal{S}_i^{\mathcal{L}} \rbrace $, we have:
\begin{align}
\langle \nabla f(\mathbf{V}_i^{\mathcal{L}}), \mathcal{P}_{\mathcal{S}_i^{\mathcal{L}}}(\mathbf{L}^\ast - \mathbf{V}_i^{\mathcal{L}}) \rangle = 0 \Rightarrow \langle \sensing \mathbf{V}_i^{\mathcal{L}} - (\obs - \sensing \mathbf{M}_i), \sensing \mathcal{P}_{\mathcal{S}_i^{\mathcal{L}}}(\mathbf{V}_i^{\mathcal{L}} - \mathbf{L}^\ast) \rangle =0. \label{eq:mALPS5:03a}
\end{align} for $ \mathcal{P}_{\mathcal{S}_i^{\mathcal{L}}}\mathbf{L}^{\ast} \in \text{span}(\mathcal{S}_i^{\mathcal{L}}) $.
Given condition (\ref{eq:mALPS5:03a}), the first term on the right hand side of (\ref{eq:mALPS5:02}) becomes: 
\begin{align}
\vectornormbig{\mathcal{P}_{\mathcal{S}_i^{\mathcal{L}}}(\mathbf{V}_i^{\mathcal{L}} - \mathbf{L}^{\ast})}_F^2 &= \langle \mathbf{V}_i^{\mathcal{L}} - \mathbf{L}^{\ast}, \mathcal{P}_{\mathcal{S}_i^{\mathcal{L}}}(\mathbf{V}_i^{\mathcal{L}} - \mathbf{L}^{\ast}) \rangle \label{eq:mALPS5:03} \\
                            &= \langle \mathbf{V}_i^{\mathcal{L}} - \mathbf{L}^{\ast}, \mathcal{P}_{\mathcal{S}_i^{\mathcal{L}}}(\mathbf{V}_i^{\mathcal{L}} - \mathbf{L}^{\ast}) \rangle - \langle \sensing \mathbf{V}_i^{\mathcal{L}} - (\obs - \sensing \mathbf{M}_i), \sensing \mathcal{P}_{\mathcal{S}_i^{\mathcal{L}}}(\mathbf{V}_i^{\mathcal{L}} - \mathbf{L}^{\ast}) \rangle \label{eq:mALPS5:04} \\ 
                            &= \langle \mathbf{V}_i^{\mathcal{L}} - \mathbf{L}^{\ast}, \mathcal{P}_{\mathcal{S}_i^{\mathcal{L}}}(\mathbf{V}_i^{\mathcal{L}} - \mathbf{L}^{\ast}) \rangle - \langle \sensing \mathbf{V}_i^{\mathcal{L}} - (\sensing (\mathbf{L}^{\ast} + \mathbf{M}^{\ast}) + \noise - \sensing \mathbf{M}_i), \sensing \mathcal{P}_{\mathcal{S}_i^{\mathcal{L}}}(\mathbf{V}_i^{\mathcal{L}} - \mathbf{L}^{\ast}) \rangle \label{eq:mALPS5:05} \\           
                            &= \langle \mathbf{V}_i^{\mathcal{L}} - \mathbf{L}^{\ast}, \mathcal{P}_{\mathcal{S}_i^{\mathcal{L}}}(\mathbf{V}_i^{\mathcal{L}} - \mathbf{L}^{\ast}) \rangle - \langle \mathbf{V}_i^{\mathcal{L}} - \mathbf{L}^{\ast} - (\mathbf{M}^{\ast} - \mathbf{M}_i), \sensing^\ast \sensing \mathcal{P}_{\mathcal{S}_i^{\mathcal{L}}}(\mathbf{V}_i^{\mathcal{L}} - \mathbf{L}^{\ast}) \rangle \nonumber \\ &+ \langle \noise, \sensing \mathcal{P}_{\mathcal{S}_i^{\mathcal{L}}}(\mathbf{V}_i^{\mathcal{L}} - \mathbf{L}^{\ast}) \rangle \label{eq:mALPS5:06} \\                            
                            &= \langle \mathbf{V}_i^{\mathcal{L}} - \mathbf{L}^{\ast}, (\id - \sensing^\ast \sensing)\mathcal{P}_{\mathcal{S}_i^{\mathcal{L}}}(\mathbf{V}_i^{\mathcal{L}} - \mathbf{L}^{\ast}) \rangle + \langle \mathbf{M}^{\ast} - \mathbf{M}_i, \sensing^\ast \sensing \mathcal{P}_{\mathcal{S}_i^{\mathcal{L}}} ( \mathbf{V}_i^{\mathcal{L}} - \mathbf{L}^{\ast})\rangle \nonumber \\ &+ \langle \noise, \sensing \mathcal{P}_{\mathcal{S}_i^{\mathcal{L}}}(\mathbf{V}_i^{\mathcal{L}} - \mathbf{L}^{\ast}) \rangle \label{eq:mALPS5:07} \\
                            &\leq | \langle \mathbf{V}_i^{\mathcal{L}} - \mathbf{L}^{\ast}, (\id - \sensing^\ast \sensing)\mathcal{P}_{\mathcal{S}_i^{\mathcal{L}}}(\mathbf{V}_i^{\mathcal{L}} - \mathbf{L}^{\ast}) \rangle | + |\langle \mathbf{M}^{\ast} - \mathbf{M}_i, \sensing^\ast \sensing \mathcal{P}_{\mathcal{S}_i^{\mathcal{L}}} ( \mathbf{V}_i^{\mathcal{L}} - \mathbf{L}^{\ast})\rangle | \nonumber \\ &+ \langle \noise, \sensing \mathcal{P}_{\mathcal{S}_i^{\mathcal{L}}}(\mathbf{V}_i^{\mathcal{L}} - \mathbf{L}^{\ast}) \rangle \label{eq:mALPS5:08}
\end{align} According to Lemma 10 in \cite{KyrillidisCevherRecipes}, we know that:
\begin{align}
| \langle \mathbf{V}_i^{\mathcal{L}} - \mathbf{L}^{\ast}, (\id - \sensing^\ast \sensing)\mathcal{P}_{\mathcal{S}_i^{\mathcal{L}}}(\mathbf{V}_i^{\mathcal{L}} - \mathbf{L}^{\ast}) \rangle | &= | \langle \mathbf{V}_i^{\mathcal{L}} - \mathbf{L}^{\ast}, (\id - \mathcal{P}_{\mathcal{S}_i^{\mathcal{L}} \cup \mathcal{L}^\ast}\sensing^\ast \sensing\mathcal{P}_{\mathcal{S}_i^{\mathcal{L}} \cup \mathcal{L}^\ast})\mathcal{P}_{\mathcal{S}_i^{\mathcal{L}}}(\mathbf{V}_i^{\mathcal{L}} - \mathbf{L}^{\ast}) \rangle | \nonumber \\ 
&\leq \vectornormbig{\mathbf{V}_i^{\mathcal{L}} - \mathbf{L}^{\ast}}_F\vectornormbig{ (\id - \mathcal{P}_{\mathcal{S}_i^{\mathcal{L}} \cup \mathcal{L}^{\ast}}\sensing^\ast \sensing\mathcal{P}_{\mathcal{S}_i^{\mathcal{L}} \cup \mathcal{L}^{\ast}})\mathcal{P}_{\mathcal{S}_i^{\mathcal{L}}}(\mathbf{V}_i^{\mathcal{L}} - \mathbf{L}^{\ast})}_F \\
&\leq \delta_{3\rank}(\sensing) \vectornormbig{\mathbf{V}_i^{\mathcal{L}} - \mathbf{L}^{\ast}}_F \vectornormbig{\mathcal{P}_{\mathcal{S}_i^{\mathcal{L}}}(\mathbf{V}_i^{\mathcal{L}} - \mathbf{L}^{\ast})}_F\label{eq:br:04}
\end{align} given the facts that $ \mathbf{V}_i^{\mathcal{L}} - \mathbf{L}^{\ast} \in \text{span}(\mathcal{S}_i^{\mathcal{L}} \cup \mathcal{L}^\ast) $ and thus $ \mathcal{P}_{\mathcal{S}_i^{\mathcal{L}} \cup \mathcal{L}^{\ast}}(\mathbf{V}_i^{\mathcal{L}} - \mathbf{L}^{\ast}) = \mathbf{V}_i^{\mathcal{L}} - \mathbf{L}^{\ast} $ and $ \mathcal{P}_{\mathcal{S}_i^{\mathcal{L}} \cup \mathcal{L}^\ast} \mathcal{P}_{\mathcal{S}_i^{\mathcal{L}}} = \mathcal{P}_{\mathcal{S}_i^{\mathcal{L}}} $ since $ \text{span}(\mathcal{S}_i^{\mathcal{L}}) \subseteq \text{span}(\mathcal{S}_i^{\mathcal{L}} \cup \mathcal{L}^\ast) $. The last inequality is due to Lemma 3 in \cite{KyrillidisCevherMatrixRecipes}. Focusing on the term $|\langle \mathbf{M}^{\ast} - \mathbf{M}_i, \sensing^\ast \sensing \mathcal{P}_{\mathcal{S}_i^{\mathcal{L}}} ( \mathbf{V}_i^{\mathcal{L}} - \mathbf{L}^{\ast})\rangle | $, we derive the following:
\begin{align}
|\langle \mathbf{M}^{\ast} - \mathbf{M}_i, \sensing^\ast \sensing \mathcal{P}_{\mathcal{S}_i^{\mathcal{L}}} ( \mathbf{V}_i^{\mathcal{L}} - \mathbf{L}^{\ast})\rangle | &= |\langle \mathcal{P}_{\mathcal{S}_i^{\mathcal{L}}} ( \mathbf{V}_i^{\mathcal{L}} - \mathbf{L}^{\ast}), \mathcal{P}_{\mathcal{S}_i^{\mathcal{L}}} \sensing \sensing (\mathbf{M}^{\ast} - \mathbf{M}_i) \rangle | \\
&\leq \vectornormbig{\mathcal{P}_{\mathcal{S}_i^{\mathcal{L}}} ( \mathbf{V}_i^{\mathcal{L}} - \mathbf{L}^{\ast})}_F \vectornormbig{\mathcal{P}_{\mathcal{S}_i^{\mathcal{L}}} \sensing \sensing (\mathbf{M}^{\ast} - \mathbf{M}_i) }_F \nonumber \\ 
&\leq \vectornormbig{\mathcal{P}_{\mathcal{S}_i^{\mathcal{L}}} ( \mathbf{V}_i^{\mathcal{L}} - \mathbf{L}^{\ast})}_F \delta_{2\rank + 2\sparsity}(\sensing) \vectornormbig{\mathbf{M}^{\ast} - \mathbf{M}_i}_F
\end{align} using Lemma 3.2 in \cite{sparcs}. Then, (\ref{eq:mALPS5:08}) becomes:
\begin{align}
\vectornormbig{\mathcal{P}_{\mathcal{S}_i^{\mathcal{L}}}(\mathbf{V}_i^{\mathcal{L}} - \mathbf{L}^{\ast})}_F^2 & \leq \delta_{3\rank}(\sensing) \vectornormbig{\mathbf{V}_i^{\mathcal{L}} - \mathbf{L}^{\ast}}_F \vectornormbig{\mathcal{P}_{\mathcal{S}_i^{\mathcal{L}}}(\mathbf{V}_i^{\mathcal{L}} - \mathbf{L}^{\ast})}_F + \delta_{2\rank + 2\sparsity}(\sensing) \vectornormbig{\mathcal{P}_{\mathcal{S}_i^{\mathcal{L}}} ( \mathbf{V}_i^{\mathcal{L}} - \mathbf{L}^{\ast})}_F \vectornormbig{\mathbf{M}^{\ast} - \mathbf{M}_i}_F \nonumber \\ &+ \sqrt{1+\delta_{2\rank}(\sensing)} \vectornormbig{\mathcal{P}_{\mathcal{S}_i^{\mathcal{L}}}(\mathbf{V}_i^{\mathcal{L}} - \mathbf{L}^{\ast})}_F \vectornormbig{\noise}_2,  \label{eq:mALPS5:09}                           
\end{align} where the last term becomes using Lemma 1 in \cite{KyrillidisCevherMatrixRecipes}. Simplifying the above quadratic expression, we obtain: 
\begin{align}
\vectornormbig{\mathcal{P}_{\mathcal{S}_i^{\mathcal{L}}}(\mathbf{V}_i^{\mathcal{L}} - \mathbf{L}^{\ast})}_F \leq \delta_{3\rank}(\sensing) \vectornormbig{\mathbf{V}_i^{\mathcal{L}} - \mathbf{L}^{\ast}}_F + \delta_{2\rank + 2\sparsity}(\sensing) \vectornormbig{\mathbf{M}^{\ast} - \mathbf{M}_i}_F + \sqrt{1+\delta_{2\rank}(\sensing)} \vectornormbig{\noise}_2. \label{eq:mALPS5:10}                           
\end{align}

As a consequence, (\ref{eq:mALPS5:02}) can be upper bounded by:
\begin{align}
\vectornormbig{\mathbf{V}_i^{\mathcal{L}} - \mathbf{L}^{\ast}}_F^2 \leq \big(\delta_{3\rank}(\sensing) \vectornormbig{\mathbf{V}_i^{\mathcal{L}} - \mathbf{L}^{\ast}}_F + \delta_{2\rank + 2\sparsity}(\sensing) \vectornormbig{\mathbf{M}^{\ast} - \mathbf{M}_i}_F  + \sqrt{1+\delta_{2\rank}(\sensing)} \vectornormbig{\noise}_2\big)^2+ \vectornormbig{\mathcal{P}_{\mathcal{S}_i^{\mathcal{L}}}^{\bot}(\mathbf{V}_i^{\mathcal{L}} - \mathbf{L}^{\ast})}_F^2. \label{eq:mALPS5:11}                           
\end{align}

We form the quadratic polynomial for this inequality assuming as unknown variable the quantity $ \vectornormbig{\mathbf{V}_i^{\mathcal{L}} - \mathbf{L}^{\ast}}_F $. Bounding by the largest root of the resulting polynomial, we get:
\begin{align}
&\vectornormbig{\mathbf{V}_i^{\mathcal{L}} - \mathbf{L}^{\ast}}_F \nonumber \leq \frac{1}{\sqrt{1-\delta_{3\rank}^2(\sensing)}} \vectornormbig{\mathcal{P}_{\mathcal{S}_i^{\mathcal{L}}}^{\bot}(\mathbf{V}_i^{\mathcal{L}} - \mathbf{L}^{\ast})}_F + \frac{(1+2\delta_{2\rank}(\sensing))}{1 - \delta_{3\rank}^2(\sensing)}\Big(\delta_{2\rank + 2\sparsity}(\sensing)\vectornormbig{\mathbf{M}^{\ast} - \mathbf{M}_i}_F + \sqrt{1+\delta_{2\rank}(\sensing)} \vectornormbig{\noise}_2\Big). 
\end{align} 

\subsection{Proof of Inequality (\ref{eq:inequality1})}

\begin{proof}
We observe the following
\begin{align}
\vectornormbig{\mathbf{L}_{i+1}  - \mathbf{L}^{\ast}}_F^2 &= \vectornormbig{\mathbf{L}_{i+1}  - \mathbf{V}_i^{\mathcal{L}} + \mathbf{V}_i^{\mathcal{L}} - \mathbf{L}^{\ast}}_F^2 \label{eq:16} \\
									   &= \vectornormbig{(\mathbf{V}_i^{\mathcal{L}} - \mathbf{L}^{\ast}) - (\mathbf{V}_i^{\mathcal{L}} - \mathbf{L}_{i+1} )}_F^2 \label{eq:17} \\
									   &= \vectornormbig{\mathbf{V}_i^{\mathcal{L}} - \mathbf{L}^{\ast}}_F^2 + \vectornormbig{\mathbf{V}_i^{\mathcal{L}} - \mathbf{L}_{i+1} }_F^2 - 2\langle \mathbf{V}_i^{\mathcal{L}} - \mathbf{L}^{\ast}, \mathbf{V}_i^{\mathcal{L}} - \mathbf{L}_{i+1}  \rangle. \label{eq:18}
\end{align}
Focusing on the right hand side of expression (\ref{eq:18}), $ \langle \mathbf{V}_i^{\mathcal{L}} - \mathbf{L}^{\ast}, \mathbf{V}_i^{\mathcal{L}} - \mathbf{L}_{i+1}  \rangle = \langle \mathbf{V}_i^{\mathcal{L}} - \mathbf{L}^{\ast}, \mathcal{P}_{\mathcal{S}_i^{\mathcal{L}}}(\mathbf{V}_i^{\mathcal{L}} - \mathbf{L}_{i+1} ) \rangle $ can be similarly analysed as in (\ref{eq:br:04}). Using the optimality condition as in (\ref{eq:mALPS5:03a}), we obtain the following expression:
\begin{align}
|\langle \mathbf{V}_i^{\mathcal{L}} - \mathbf{L}^{\ast}, \mathcal{P}_{\mathcal{S}_i^{\mathcal{L}}}(\mathbf{V}_i^{\mathcal{L}} - \mathbf{L}_{i+1} ) \rangle | 
&= |\langle \mathbf{V}_i^{\mathcal{L}} - \mathbf{L}^{\ast}, \mathcal{P}_{\mathcal{S}_i^{\mathcal{L}}}(\mathbf{V}_i^{\mathcal{L}} - \mathbf{L}_{i+1} ) \rangle - \langle \mathbf{V}_i^{\mathcal{L}} - \mathbf{L}^{\ast} - (\mathbf{M}^{\ast} - \mathbf{M}_i), \sensing^\ast \sensing \mathcal{P}_{\mathcal{S}_i^{\mathcal{L}}}(\mathbf{V}_i^{\mathcal{L}} - \mathbf{L}_{i+1}) \rangle \nonumber \\ &+ \langle \noise, \sensing \mathcal{P}_{\mathcal{S}_i^{\mathcal{L}}}(\mathbf{V}_i^{\mathcal{L}} - \mathbf{L}_{i+1}) \rangle | \nonumber \\ 
&= |\langle \mathbf{V}_i^{\mathcal{L}} - \mathbf{L}^{\ast}, (I - \sensing^\ast\sensing)\mathcal{P}_{\mathcal{S}_i^{\mathcal{L}}}(\mathbf{V}_i^{\mathcal{L}} - \mathbf{L}_{i+1} ) \rangle + \langle \mathbf{M}^{\ast} - \mathbf{M}_i, \sensing^\ast \sensing \mathcal{P}_{\mathcal{S}_i^{\mathcal{L}}}(\mathbf{V}_i^{\mathcal{L}} - \mathbf{L}_{i+1}) \rangle \nonumber \\ &+ \langle \noise, \sensing \mathcal{P}_{\mathcal{S}_i^{\mathcal{L}}}(\mathbf{V}_i^{\mathcal{L}} - \mathbf{L}_{i+1}) \rangle | \nonumber \\ 
&\leq \delta_{3\rank}(\sensing) \vectornormbig{\mathbf{V}_i^{\mathcal{L}} - \mathbf{L}^{\ast}}_F \vectornormbig{\mathbf{V}_i^{\mathcal{L}} - \mathbf{L}_{i+1} }_F  + \vectornormbig{\mathcal{P}_{\mathcal{S}_i^{\mathcal{L}}} \sensing^{\ast} \sensing (\mathbf{M}^{\ast} - \mathbf{M}_i)}_F \vectornormbig{\mathbf{V}_i^{\mathcal{L}} - \mathbf{L}_{i+1}}_F \nonumber \\ &+ \sqrt{1 +\delta_{2\rank}(\sensing)} \vectornormbig{\noise}_2 \vectornormbig{\mathbf{V}_i^{\mathcal{L}} - \mathbf{L}_{i+1}}_F \nonumber \\ 
&\leq \delta_{3\rank}(\sensing) \vectornormbig{\mathbf{V}_i^{\mathcal{L}} - \mathbf{L}^{\ast}}_F \vectornormbig{\mathbf{V}_i^{\mathcal{L}} - \mathbf{L}_{i+1} }_F  + \delta_{2\rank + 2\sparsity}(\sensing)\vectornormbig{\mathbf{M}^{\ast} - \mathbf{M}_i}_F \vectornormbig{\mathbf{V}_i^{\mathcal{L}} - \mathbf{L}_{i+1}}_F \nonumber \\ &+ \sqrt{1 +\delta_{2\rank}(\sensing)} \vectornormbig{\noise}_2 \vectornormbig{\mathbf{V}_i^{\mathcal{L}} - \mathbf{L}_{i+1}}_F. \label{eq:24}
\end{align}

Now, expression (\ref{eq:18}) can be further transformed as:
\begin{align}
\vectornormbig{\mathbf{L}_{i+1}  - \mathbf{L}^{\ast}}_F^2 &= \vectornormbig{\mathbf{V}_i^{\mathcal{L}} - \mathbf{L}^{\ast}}_F^2 + \vectornormbig{\mathbf{V}_i^{\mathcal{L}} - \mathbf{L}_{i+1} }_F^2 - 2\langle \mathbf{V}_i^{\mathcal{L}} - \mathbf{L}^{\ast}, \mathbf{V}_i^{\mathcal{L}} - \mathbf{L}_{i+1}  \rangle \label{eq:25} \\	
										&\leq \vectornormbig{\mathbf{V}_i^{\mathcal{L}} - \mathbf{L}^{\ast}}_F^2 + \vectornormbig{\mathbf{V}_i^{\mathcal{L}} - \mathbf{L}_{i+1} }_F^2 + 2|\langle \mathbf{V}_i^{\mathcal{L}} - \mathbf{L}^{\ast}, \mathbf{V}_i^{\mathcal{L}} - \mathbf{L}_{i+1}  \rangle |\label{eq:25a} \\	
										&\stackrel{(i)}{\leq} \vectornormbig{\mathbf{V}_i^{\mathcal{L}} - \mathbf{L}^{\ast}}_F^2 + \vectornormbig{\mathbf{V}_i^{\mathcal{L}} - \mathbf{L}_{i+1} }_F^2 +  2(\delta_{3\rank}(\sensing) \vectornormbig{\mathbf{V}_i^{\mathcal{L}} - \mathbf{L}^{\ast}}_F \vectornormbig{\mathbf{V}_i^{\mathcal{L}} - \mathbf{L}_{i+1} }_F  \nonumber \\ &+ \delta_{2\rank + 2\sparsity}(\sensing)\vectornormbig{\mathbf{M}^{\ast} - \mathbf{M}_i}_F \vectornormbig{\mathbf{V}_i^{\mathcal{L}} - \mathbf{L}_{i+1}}_F + \sqrt{1 +\delta_{2\rank}(\sensing)} \vectornormbig{\noise}_2 \vectornormbig{\mathbf{V}_i^{\mathcal{L}} - \mathbf{L}_{i+1}}_F.) \label{eq:25b}										                                      
\end{align} where $ (i) $ is due to (\ref{eq:24}). Using the inequality $ \vectornorm{\mathbf{L}_{i+1} - \mathbf{V}_i^{\mathcal{L}}}_F \leq \vectornormbig{\mathcal{P}_{\mathcal{S}_i^{\mathcal{L}}}(\mathbf{V}_i^{\mathcal{L}} - \mathbf{L}^{\ast})}_F $, we get:
\begin{align}
\vectornormbig{\mathbf{L}_{i+1}  - \mathbf{L}^{\ast}}_F^2 &\leq \vectornormbig{\mathbf{V}_i^{\mathcal{L}} - \mathbf{L}^{\ast}}_F^2 + \vectornormbig{\mathcal{P}_{\mathcal{S}_i^{\mathcal{L}}}(\mathbf{V}_i^{\mathcal{L}} - \mathbf{L}^{\ast})}_F ^2 +  2(\delta_{3\rank}(\sensing) \vectornormbig{\mathbf{V}_i^{\mathcal{L}} - \mathbf{L}^{\ast}}_F \vectornormbig{\mathcal{P}_{\mathcal{S}_i^{\mathcal{L}}}(\mathbf{V}_i^{\mathcal{L}} - \mathbf{L}^{\ast})}_F  \nonumber \\ &+ \delta_{2\rank + 2\sparsity}(\sensing)\vectornormbig{\mathbf{M}^{\ast} - \mathbf{M}_i}_F \vectornormbig{\mathcal{P}_{\mathcal{S}_i^{\mathcal{L}}}(\mathbf{V}_i^{\mathcal{L}} - \mathbf{L}^{\ast})}_F + \sqrt{1 +\delta_{2\rank}(\sensing)} \vectornormbig{\noise}_2 \vectornormbig{\mathcal{P}_{\mathcal{S}_i^{\mathcal{L}}}(\mathbf{V}_i^{\mathcal{L}} - \mathbf{L}^{\ast})}_F.)  \label{eq:mALPS5:20}
\end{align} Furthermore, replacing $ \vectornormbig{\mathcal{P}_{\mathcal{S}_i^{\mathcal{L}}}(\mathbf{L}^{\ast} - \mathbf{V}_i^{\mathcal{L}})}_F $ with its upper bound defined in (\ref{eq:mALPS5:10}), we compute:
\begin{align}
\vectornormbig{\mathbf{L}_{i+1}  - \mathbf{L}^{\ast}}_F^2 &=\big(1 + 3\delta_{3\rank}^2(\sensing)\big) \vectornormbig{\mathbf{V}_i^{\mathcal{L}} - \mathbf{L}^{\ast}}_F^2 + 4\delta_{3\rank}(\sensing)\vectornormbig{\mathbf{V}_i^{\mathcal{L}} - \mathbf{L}^{\ast}}_F(\delta_{2\rank + 2\sparsity}(\sensing) \vectornormbig{\mathbf{M}^{\ast} - \mathbf{M}_i}_F + \sqrt{1+\delta_{2\rank}(\sensing)} \vectornormbig{\noise}_2) \nonumber \\ &+ 3(\sqrt{1+\delta_{2\rank}(\sensing)}\vectornormbig{\noise}_2 + \delta_{2\rank + 2\sparsity}(\sensing)\vectornormbig{\mathbf{M}^{\ast} - \mathbf{M}_i}_F)^2  \nonumber \\
															   &\stackrel{(i)}{\leq} \Big(1 + 3\delta_{3\rank}^2(\sensing)\Big)\Bigg(\vectornormbig{\mathbf{V}_i^{\mathcal{L}} - \mathbf{L}^{\ast}}_F + \sqrt{\frac{3}{1+ 3\delta_{3\rank}^2(\sensing)}} \big(\delta_{2\rank + 2\sensing}(\sensing) \vectornormbig{\mathbf{M}^{\ast} - \mathbf{M}_i}_F + \sqrt{1+\delta_{2\rank}(\sensing)} \vectornormbig{\noise}\big)\Bigg)^2 \nonumber
\end{align} where $ (i) $ is obtained by completing the squares and eliminating negative terms. Thus:
\begin{align}
\vectornormbig{\mathbf{L}_{i+1}  - \mathbf{L}^{\ast}}_F \leq \sqrt{1 + 3\delta_{3\rank}^2(\sensing)}\Bigg(\vectornormbig{\mathbf{V}_i^{\mathcal{L}} - \mathbf{L}^{\ast}}_F + \sqrt{\frac{3}{1+ 3\delta_{3\rank}^2(\sensing)}} \big(\delta_{2\rank + 2\sensing}(\sensing) \vectornormbig{\mathbf{M}^{\ast} - \mathbf{M}_i}_F + \sqrt{1+\delta_{2\rank}(\sensing)} \vectornormbig{\noise}\big)\Bigg) \nonumber
\end{align} Furthermore, we exploit Lemma \ref{lemma:leastsquares_lowrank} to obtain inequality (\ref{eq:inequality1}).
\end{proof}

\subsection{Proof of Theorem \ref{theorem:2}}
Here, we prove the convergence of Algorithm 2, both for the low rank and the sparse matrix estimate part, and then combine the corresponding theoretical results. Let $ \mathcal{L}^\ast \leftarrow \text{ortho}(\mathbf{L}^{\ast}) $ be a set of orthonormal, rank-1 matrices that span the range of $ \mathbf{L}^{\ast} $ and $\mathcal{M}^{\ast} $ be the set of indices of the non-zero elemetns in $\mathbf{M}^\ast$. For the low rank matrix estimate, we observe the following:
\begin{align}
\vectornormbig{\mathbf{L}_{i+1} - \mathbf{V}_i^{\mathcal{L}}}_F^2 &\leq \vectornormbig{\mathbf{L}^{\ast} - \mathbf{V}_i^{\mathcal{L}}}_F^2 \Rightarrow \\
\vectornormbig{\mathbf{L}_{i+1} - \mathbf{L}^{\ast} + \mathbf{L}^{\ast} - \mathbf{V}_i^{\mathcal{L}}}_F^2 &\leq \vectornormbig{\mathbf{L}^{\ast} - \mathbf{V}_i^{\mathcal{L}}}_F^2 \Rightarrow \\
\vectornormbig{\mathbf{L}_{i+1} - \mathbf{L}^{\ast}}_F^2 + \vectornormbig{\mathbf{V}_i^{\mathcal{L}} - \mathbf{L}^{\ast}}_F^2 + 2\langle \mathbf{L}_{i+1} - \mathbf{L}^{\ast}, \mathbf{L}^{\ast} - \mathbf{V}_i^{\mathcal{L}} \rangle &\leq \vectornormbig{\mathbf{L}^{\ast} - \mathbf{V}_i^{\mathcal{L}}}_F^2 \Rightarrow \\
\vectornormbig{\mathbf{L}_{i+1} - \mathbf{L}^{\ast}}_F^2 &\leq 2\langle \mathbf{L}_{i+1} - \mathbf{L}^{\ast}, \mathbf{V}_i^{\mathcal{L}} - \mathbf{L}^{\ast} \rangle \label{eq:mALPS0_memory:00}
\end{align}

From Algorithm 2, it is obvious that $ (i) ~\mathbf{V}_i^{\mathcal{L}} \in \text{span}(\mathcal{S}_i^{\mathcal{L}}) $, $ (ii) $ $ \mathbf{Q}_i^{\mathcal{L}}\in \text{span}( \mathcal{S}_i^{\mathcal{L}}) $ and $ (iii) $ $ \mathbf{L}_{i+1} \in \text{span}(\mathcal{S}_i^{\mathcal{L}}) $. We define $ \mathcal{E} := \mathcal{S}_i^{\mathcal{L}} \cup \mathcal{L}^{\ast} $ where $ \text{rank}(\text{span}(\mathcal{E})) \leq 4\rank $ and let $ \mathcal{P}_{\mathcal{E}} $ be the orthogonal projection onto the subspace defined by $ \mathcal{E} $. We highlight that $\mathcal{P}_{\mathcal{E}} \mathcal{P}_{\mathcal{S}_i^{\mathcal{L}}} = \mathcal{P}_{\mathcal{S}_i^{\mathcal{L}}}$.

Since $ \mathbf{L}_{i+1} - \mathbf{L}^{\ast} \in \text{span}(\mathcal{E}) $ and $ \mathbf{V}_i^{\mathcal{L}} - \mathbf{L}^{\ast} \in \text{span}(\mathcal{E}) $, the following hold true:
\begin{align}
\mathbf{L}_{i+1} - \mathbf{L}^{\ast} = \mathcal{P}_{\mathcal{E}} (\mathbf{L}_{i+1} - \mathbf{L}^{\ast})~~\text{and}~~\mathbf{V}_i^{\mathcal{L}} - \mathbf{L}^{\ast} = \mathcal{P}_{\mathcal{E}} (\mathbf{V}_i^{\mathcal{L}} - \mathbf{L}^{\ast}).
\end{align}

Then, (\ref{eq:mALPS0_memory:00}) can be written as:
\begin{align}
&\vectornormbig{\mathbf{L}_{i+1} - \mathbf{L}^{\ast}}_F^2 \\
&\leq 2\langle \mathcal{P}_{\mathcal{E}}(\mathbf{L}_{i+1} - \mathbf{L}^{\ast}), \mathcal{P}_{\mathcal{E}}(\mathbf{V}_i^{\mathcal{L}} - \mathbf{L}^{\ast}) \rangle \\
 &= 2\langle \mathcal{P}_{\mathcal{E}}(\mathbf{L}_{i+1} - \mathbf{L}^{\ast}), \mathcal{P}_{\mathcal{E}}\left(\mathbf{Q}_i^{\mathcal{L}}+ \mu_i^{\mathcal{L}} \mathcal{P}_{\mathcal{S}_i^{\mathcal{L}}} \sensing^\ast (\obs - \sensing \mathbf{Q}_i) - \mathbf{L}^{\ast}\right) \rangle \\
 &= 2\langle \mathbf{L}_{i+1} - \mathbf{L}^{\ast}, \mathcal{P}_{\mathcal{E}}(\mathbf{Q}_i^{\mathcal{L}} - \mathbf{L}^{\ast}) + \mu_i^{\mathcal{L}} \mathcal{P}_{\mathcal{E}} \mathcal{P}_{\mathcal{S}_i^{\mathcal{L}}}\big(\sensing^\ast (\sensing (\mathbf{L}^{\ast} + \mathbf{M}^{\ast}) - \sensing \mathbf{Q}_i)\big) \rangle \label{eq:mALPS0_memory:01} \\
&= 2\langle \mathbf{L}_{i+1} - \mathbf{L}^{\ast}, \mathcal{P}_{\mathcal{E}}(\mathbf{Q}_i^{\mathcal{L}} - \mathbf{L}^{\ast}) + \mu_i^{\mathcal{L}} \mathcal{P}_{\mathcal{E}} \mathcal{P}_{\mathcal{S}_i^{\mathcal{L}}}\big(\sensing^\ast \sensing (\mathbf{L}^{\ast} + \mathbf{M}^{\ast}) - \sensing^{\ast} \sensing (\mathbf{Q}_i^{\mathcal{L}} + \mathbf{Q}_i^{\mathcal{M}}\big) \rangle  \\
&= 2\langle \mathbf{L}_{i+1} - \mathbf{L}^{\ast}, \mathcal{P}_{\mathcal{E}}(\mathbf{Q}_i^{\mathcal{L}} - \mathbf{L}^{\ast}) - \mu_i^{\mathcal{L}} \mathcal{P}_{\mathcal{E}} \mathcal{P}_{\mathcal{S}_i^{\mathcal{L}}} \sensing^\ast \sensing (\mathbf{Q}_i^{\mathcal{L}} - \mathbf{L}^{\ast}) - \mu_i^{\mathcal{L}} \mathcal{P}_{\mathcal{E}} \mathcal{P}_{\mathcal{S}_i^{\mathcal{L}}} \sensing^{\ast} \sensing (\mathbf{Q}_i^{\mathcal{M}} - \mathbf{M}^{\ast}) \rangle  \\
&= 2\langle \mathbf{L}_{i+1} - \mathbf{L}^{\ast}, \mathcal{P}_{\mathcal{E}}(\mathbf{Q}_i^{\mathcal{L}} - \mathbf{L}^{\ast}) - \mu_i^{\mathcal{L}} \mathcal{P}_{\mathcal{E}} \mathcal{P}_{\mathcal{S}_i^{\mathcal{L}}} \sensing^\ast \sensing \mathcal{P}_{\mathcal{E}}(\mathbf{Q}_i^{\mathcal{L}} - \mathbf{L}^{\ast})\rangle - 2\mu_i^{\mathcal{L}} \langle \mathbf{L}_{i+1} - \mathbf{L}^{\ast}, \mathcal{P}_{\mathcal{E}} \mathcal{P}_{\mathcal{S}_i^{\mathcal{L}}} \sensing^{\ast} \sensing (\mathbf{Q}_i^{\mathcal{M}} - \mathbf{M}^{\ast}) \rangle \\ 
&= 2\langle \mathbf{L}_{i+1} - \mathbf{L}^{\ast}, \mathcal{P}_{\mathcal{E}}(\mathbf{Q}_i^{\mathcal{L}} - \mathbf{L}^{\ast}) - \mu_i^{\mathcal{L}} \mathcal{P}_{\mathcal{E}} \mathcal{P}_{\mathcal{S}_i^{\mathcal{L}}} \sensing^\ast \sensing \big[\mathcal{P}_{\mathcal{S}_i^{\mathcal{L}}} + \mathcal{P}_{\mathcal{S}_i^{\mathcal{L}}}^{\bot}\big]\mathcal{P}_{\mathcal{E}}(\mathbf{Q}_i^{\mathcal{L}} - \mathbf{L}^{\ast})\rangle \nonumber \\ &- 2\mu_i^{\mathcal{L}} \langle \mathbf{L}_{i+1} - \mathbf{L}^{\ast}, \mathcal{P}_{\mathcal{E}} \mathcal{P}_{\mathcal{S}_i^{\mathcal{L}}} \sensing^{\ast} \sensing (\mathbf{Q}_i^{\mathcal{M}} - \mathbf{M}^{\ast}) \rangle \label{koutourou:1}
\end{align} due to $ \mathcal{P}_{\mathcal{E}}(\mathbf{Q}_i^{\mathcal{L}}- \mathbf{L}^{\ast}) := \mathcal{P}_{\mathcal{S}_i^{\mathcal{L}}} \mathcal{P}_{\mathcal{E}}(\mathbf{Q}_i^{\mathcal{L}}- \mathbf{L}^{\ast}) + \mathcal{P}_{\mathcal{S}_i^{\mathcal{L}}}^{\bot} \mathcal{P}_{\mathcal{E}}(\mathbf{Q}_i^{\mathcal{L}}- \mathbf{L}^{\ast}) $. The first term in (\ref{koutourou:1}) satsifies:
\begin{align}
&2\langle \mathbf{L}_{i+1} - \mathbf{L}^{\ast}, \mathcal{P}_{\mathcal{E}}(\mathbf{Q}_i^{\mathcal{L}} - \mathbf{L}^{\ast}) - \mu_i^{\mathcal{L}} \mathcal{P}_{\mathcal{E}} \mathcal{P}_{\mathcal{S}_i^{\mathcal{L}}} \sensing^\ast \sensing \big[\mathcal{P}_{\mathcal{S}_i^{\mathcal{L}}} + \mathcal{P}_{\mathcal{S}_i^{\mathcal{L}}}^{\bot}\big]\mathcal{P}_{\mathcal{E}}(\mathbf{Q}_i^{\mathcal{L}} - \mathbf{L}^{\ast})\rangle \nonumber \\ 
&\leq 2 \vectornormbig{\mathbf{L}_{i+1} - \mathbf{L}^{\ast}}_F \vectornormbig{(I - \mu_i^{\mathcal{L}} \mathcal{P}_{\mathcal{E}} \mathcal{P}_{\mathcal{S}_i^{\mathcal{L}}} \sensing^\ast \sensing \mathcal{P}_{\mathcal{S}_i^{\mathcal{L}}}) \mathcal{P}_{\mathcal{E}}(\mathbf{Q}_i^{\mathcal{L}} - \mathbf{L}^{\ast})}_F + 2\mu_i^{\mathcal{L}} \vectornormbig{\mathbf{L}_{i+1} - \mathbf{L}^{\ast}}_F  \vectornormbig{\mathcal{P}_{\mathcal{S}_i^{\mathcal{L}}} \sensing^\ast \sensing \mathcal{P}_{\mathcal{S}_i^{\mathcal{L}}}^\bot \mathcal{P}_{\mathcal{E}}(\mathbf{Q}_i^{\mathcal{L}} - \mathbf{L}^{\ast})}_F \nonumber \\
&\leq \frac{4 \delta_{3\rank}(\sensing)}{1 - \delta_{3\rank}(\sensing)} \vectornormbig{\mathbf{L}_{i+1} - \mathbf{L}^{\ast}}_F \vectornormbig{\mathbf{Q}_i^{\mathcal{L}} - \mathbf{L}^{\ast}}_F + \frac{2\delta_{4\rank}(\sensing)}{1- \delta_{3\rank}(\sensing)}(2\delta_{3\rank}(\sensing) + 2\delta_{4\rank}(\sensing)) \vectornormbig{\mathbf{L}_{i+1} - \mathbf{L}^{\ast}}_F \vectornormbig{\mathbf{Q}_i^{\mathcal{L}} - \mathbf{L}^{\ast}}_F \label{koutourou:2}
\end{align} where (\ref{koutourou:2}) holds, since $ \frac{1}{1+\delta_{3\rank}(\sensing)} \leq \mu_i^{\mathcal{L}} \leq \frac{1}{1-\delta_{3\rank}(\sensing)} $, using Lemma 3 in \cite{KyrillidisCevherMatrixRecipes}:
\begin{align}
\lambda(\id - \mu_i^{\mathcal{L}}\mathcal{P}_{\mathcal{S}_i^{\mathcal{L}}}\sensing^\ast \sensing \mathcal{P}_{\mathcal{S}_i^{\mathcal{L}}}) \in \Bigg[1 - \frac{1-\delta_{3\rank}(\sensing)}{1+\delta_{3\rank}(\sensing)}, \frac{1+\delta_{3\rank}(\sensing)}{1-\delta_{3\rank}(\sensing)} - 1\Bigg] \leq \frac{2\delta_{3\rank}(\sensing)}{1-\delta_{3\rank}(\sensing)}.
\end{align} and thus:
\begin{align}
\vectornormbig{(\id - \mu_i^{\mathcal{L}}\mathcal{P}_{\mathcal{S}_i^{\mathcal{L}}}\sensing^\ast \sensing \mathcal{P}_{\mathcal{S}_i^{\mathcal{L}}})\mathcal{P}_{\mathcal{E}}(\mathbf{Q}_i^{\mathcal{L}}- \mathbf{L}^{\ast})}_F \leq \frac{2\delta_{3\rank}(\sensing)}{1-\delta_{3\rank}(\sensing)} \vectornormbig{\mathcal{P}_{\mathcal{E}}(\mathbf{Q}_i^{\mathcal{L}}- \mathbf{L}^{\ast})}_F.
\end{align} Furthermore, according to Lemma 4 in \cite{KyrillidisCevherMatrixRecipes}:
\begin{align}
\vectornormbig{\mathcal{P}_{\mathcal{S}_i^{\mathcal{L}}}\sensing^\ast \sensing \mathcal{P}_{\mathcal{S}_i^{\mathcal{L}}}^{\bot}\mathcal{P}_{\mathcal{E}}(\mathbf{Q}_i^{\mathcal{L}}- \mathbf{L}^{\ast})}_F \leq  \delta_{4\rank}(\sensing) \vectornormbig{\mathcal{P}_{\mathcal{S}_i^{\mathcal{L}}}^{\bot}\mathcal{P}_{\mathcal{E}}(\mathbf{Q}_i^{\mathcal{L}}- \mathbf{L}^{\ast})}_F
\end{align} since $ \text{rank}(\mathcal{P}_{\mathcal{E} \cup \mathcal{S}_i^{\mathcal{L}}}\boldsymbol{Q}) \leq 4\rank, ~\forall \boldsymbol{Q} \in \mathcal{R}^{\dimension} $. Moreover:
\begin{align}
\vectornormbig{\mathcal{P}_{\mathcal{S}_i^{\mathcal{L}}}^{\bot}\mathcal{P}_{\mathcal{E}}(\mathbf{Q}_i^{\mathcal{L}}- \mathbf{L}^{\ast})}_F \leq (2\delta_{3\rank}(\sensing) + 2\delta_{4\rank}(\sensing))\vectornormbig{\mathbf{Q}_i^{\mathcal{L}}- \mathbf{L}^{\ast}}_F,
\end{align} using ideas from Lemma \ref{lemma:act_subspace_exp}. 

The second term in (\ref{koutourou:1}) satsifies:
\begin{align}
2\mu_i^{\mathcal{L}} \langle \mathbf{L}_{i+1} - \mathbf{L}^{\ast}, \mathcal{P}_{\mathcal{E}} \mathcal{P}_{\mathcal{S}_i^{\mathcal{L}}} \sensing^{\ast} \sensing (\mathbf{Q}_i^{\mathcal{M}} - \mathbf{M}^{\ast}) \rangle &\leq \frac{2}{1-\delta_{3\rank}(\sensing)} \vectornormbig{\mathbf{L}_{i+1} - \mathbf{L}^{\ast}}_F \vectornormbig{\mathcal{P}_{\mathcal{S}_i^{\mathcal{L}}} \sensing^\ast \sensing (\mathbf{Q}_i^{\mathcal{M}} - \mathbf{M}^{\ast})}_F \nonumber \\
&\leq \frac{2}{1-\delta_{3\rank}(\sensing)} \vectornormbig{\mathbf{L}_{i+1} - \mathbf{L}^{\ast}}_F \delta_{3\rank + 3\sparsity}(\sensing)\vectornormbig{ \mathbf{Q}_i^{\mathcal{M}} - \mathbf{M}^{\ast}}_F \nonumber
\end{align} using Lemma 3.2 in \cite{sparcs}. Replacing the above results in (\ref{koutourou:1}), we compute:
\begin{align}
\vectornormbig{\mathbf{L}_{i+1} - \mathbf{L}^{\ast}}_F &\leq \alpha \vectornormbig{\mathbf{Q}_i^{\mathcal{L}} - \mathbf{L}^{\ast}}_F + \beta\vectornormbig{ \mathbf{Q}_i^{\mathcal{M}} - \mathbf{M}^{\ast}}_F, \label{lowrank_part}
\end{align} where $\alpha := \Big(\frac{4 \delta_{3\rank}(\sensing)}{1 - \delta_{3\rank}(\sensing)} + \frac{2\delta_{4\rank}(\sensing)}{1- \delta_{3\rank}(\sensing)}(2\delta_{3\rank}(\sensing) + 2\delta_{4\rank}(\sensing))\Big)$ and $\beta := \frac{2\delta_{3\rank + 3\sparsity}(\sensing)}{1-\delta_{3\rank}(\sensing)} $. Following similar steps for the sparse matrix estimate part, we end up with the following inequality bound for $\mathbf{M}_{i+1}$:
\begin{align}
\vectornormbig{\mathbf{M}_{i+1} - \mathbf{M}^{\ast}}_F &\leq \gamma \vectornormbig{\mathbf{Q}_i^{\mathcal{M}} - \mathbf{M}^{\ast}}_F + \zeta \vectornormbig{ \mathbf{Q}_i^{\mathcal{L}} - \mathbf{L}^{\ast}}_F, \label{sparse_part}
\end{align} where $\gamma := \frac{2(\delta_{4\sparsity}(\sensing) + \delta_{3\sparsity}(\sensing))}{1- \delta_{3\sparsity}(\sensing)} $ and $\zeta := \frac{2\delta_{3\rank + 4\sparsity}(\sensing)}{1-\delta_{3\sparsity}(\sensing)} $.

Furthermore:
\begin{align}
\vectornormbig{\mathbf{Q}_i^{\mathcal{L}}- \mathbf{L}^{\ast}}_F &= \vectornormbig{\mathbf{L}_i + \tau_i(\mathbf{L}_i - \mathbf{L}_{i-1}) - \mathbf{L}^{\ast}}_F \nonumber \\ &= \vectornormbig{(1+\tau_i)(\mathbf{L}_i - \mathbf{L}^{\ast}) + \tau_i(\mathbf{L}^{\ast} - \mathbf{L}_{i-1})}_F \nonumber \\ &\leq (1+\tau_i)\vectornormbig{\mathbf{L}_i - \mathbf{L}^{\ast}}_F + \tau_i\vectornormbig{\mathbf{L}_{i-1} - \mathbf{L}^{\ast}}_F \label{eq:mALPS_memory:05}
\end{align} and 
\begin{align}
\vectornormbig{\mathbf{Q}_i^{\mathcal{M}}- \mathbf{M}^{\ast}}_F &= \vectornormbig{\mathbf{M}_i + \tau_i(\mathbf{M}_i - \mathbf{M}_{i-1}) - \mathbf{M}^{\ast}}_F \nonumber \\ &= \vectornormbig{(1+\tau_i)(\mathbf{M}_i - \mathbf{M}^{\ast}) + \tau_i(\mathbf{M}^{\ast} - \mathbf{M}_{i-1})}_F \nonumber \\ &\leq (1+\tau_i)\vectornormbig{\mathbf{M}_i - \mathbf{M}^{\ast}}_F + \tau_i\vectornormbig{\mathbf{M}_{i-1} - \mathbf{M}^{\ast}}_F \label{eq:mALPS_memory:05b}
\end{align} Combining (\ref{eq:mALPS_memory:05}), (\ref{eq:mALPS_memory:05b}) into (\ref{lowrank_part}) and (\ref{sparse_part}), we get:
\begin{align}
\vectornormbig{\mathbf{L}_{i+1} - \mathbf{L}^{\ast}}_F &\leq \alpha (1+\tau_i) \vectornormbig{\mathbf{L}_{i} - \mathbf{L}^{\ast}}_F + \alpha \tau_i \vectornormbig{\mathbf{L}_{i-1} - \mathbf{L}^{\ast}}_F \nonumber \\ & + \beta (1+\tau_i) \vectornormbig{\mathbf{M}_{i} - \mathbf{M}^{\ast}}_F + \beta \tau_i \vectornormbig{\mathbf{M}_{i-1} - \mathbf{M}^{\ast}}_F \label{lowrank_part_2}
\end{align} and
\begin{align}
\vectornormbig{\mathbf{M}_{i+1} - \mathbf{M}^{\ast}}_F &\leq \gamma (1+\tau_i) \vectornormbig{\mathbf{M}_{i} - \mathbf{M}^{\ast}}_F + \gamma \tau_i \vectornormbig{\mathbf{M}_{i-1} - \mathbf{M}^{\ast}}_F \nonumber \\ & + \zeta (1+\tau_i) \vectornormbig{\mathbf{L}_{i} - \mathbf{L}^{\ast}}_F + \zeta \tau_i \vectornormbig{\mathbf{L}_{i-1} - \mathbf{L}^{\ast}}_F \label{sparse_part_2}
\end{align}
The inequalities (\ref{lowrank_part_2}) and (\ref{sparse_part_2}) define the following coupled set of inequalities:
\begin{align}
\begin{bmatrix}
\vectornormbig{\mathbf{L}_{i+1} - \mathbf{L}^{\ast}}_F \\ \vectornormbig{\mathbf{M}_{i+1} - \mathbf{M}^{\ast}}_F
\end{bmatrix} \leq (1+\tau_i) \boldsymbol{\Delta} \begin{bmatrix}
\vectornormbig{\mathbf{L}_{i} - \mathbf{L}^{\ast}}_F \\ \vectornormbig{\mathbf{M}_{i} - \mathbf{M}^{\ast}}_F \end{bmatrix} + \tau_i \boldsymbol{\Delta} \begin{bmatrix}
\vectornormbig{\mathbf{L}_{i-1} - \mathbf{L}^{\ast}}_F \\ \vectornormbig{\mathbf{M}_{i-1} - \mathbf{M}^{\ast}}_F \label{memory_ineq_1}
\end{bmatrix}
\end{align} where $\boldsymbol{\Delta} := \begin{bmatrix} \alpha~~ \beta \\ \zeta ~~\gamma \end{bmatrix}$. Furthermore, we define $\mathbf{x}(i) := \begin{bmatrix} \vectornormbig{\mathbf{L}_i - \mathbf{L}^{\ast}}_F \\ \vectornormbig{\mathbf{M}_i - \mathbf{M}^{\ast}}_F \end{bmatrix}$ to obtain inequality (\ref{memory_original_inequality}). We can convert this second-order linear system into a two-dimensional first-order system where the variables of the linear system are multi-dimensional. To achieve this, we define a new state variable $\mathbf{y}(i)$ where:
\begin{align}
\mathbf{y}(i) := \mathbf{x}(i+1).
\end{align} and thus, $ \mathbf{y}(i+1) := \mathbf{x}(i+2) $. Using the new variable above, we define the following two-dimensional first-order system:
\begin{align}
\left\{
	\begin{array}{ll}
		\mathbf{y}(i+1) - (1+\tau_i)\boldsymbol{\Delta} \mathbf{y}(i) - \tau_i \boldsymbol{\Delta} \mathbf{x}(i) \leq 0, \\
		\mathbf{x}(i+1) \leq \mathbf{y}(i).
	\end{array}
\right.
\end{align} which, moreover, defines the following linear system that characterizes the evolution of two state variables, $ \lbrace \mathbf{y}(i), \mathbf{x}(i) \rbrace $:
\begin{align}
\begin{bmatrix}
\mathbf{y}(i+1) \\ \mathbf{x}(i+1)
\end{bmatrix} \leq \begin{bmatrix} 
(1+\tau_i)\boldsymbol{\Delta} & \tau_i \boldsymbol{\Delta} \\
\mathbb{I} &\mathbf{0}
\end{bmatrix} \begin{bmatrix} \mathbf{y}(i) \\ \mathbf{x}(i) \end{bmatrix} \Rightarrow \begin{bmatrix}
\mathbf{x}(i+2) \\ \mathbf{x}(i+1)
\end{bmatrix} \leq \begin{bmatrix} 
(1+\tau_i)\boldsymbol{\Delta} & \tau_i \boldsymbol{\Delta} \\
\mathbb{I} &\mathbf{0}
\end{bmatrix} \begin{bmatrix} \mathbf{x}(i+1) \\ \mathbf{x}(i) \end{bmatrix},
\end{align} with well-defined initial conditions $\mathbf{x}(0) := \begin{bmatrix} \vectornormbig{\mathbf{L}^{\ast}}_F \\ \vectornormbig{\mathbf{M}^{\ast}}_F \end{bmatrix} $ and $\mathbf{y}(0) := \mathbf{x}(1) = (1+\tau_i)\boldsymbol{\Delta}\mathbf{x}(0) $. For $\mathbf{w}(i) := \begin{bmatrix} \mathbf{x}(i+1) \\ \mathbf{x}(i) \end{bmatrix} $, we obtain the linear system:
\begin{align} 
\mathbf{w}(i+1) \leq \underbrace{\begin{bmatrix} (1+\tau_i)\boldsymbol{\Delta} & \tau_i \boldsymbol{\Delta} \\ \mathbb{I} & \mathbf{0} \end{bmatrix}}_{\widehat{\boldsymbol{\Delta}}} \mathbf{w}(i).
\end{align}
Unfolding the recursion, we get the inequality (\ref{recursive}):
\begin{align} 
\mathbf{w}(i+1) \leq \widehat{\boldsymbol{\Delta}}^{i} \mathbf{w}(0).
\end{align} Assuming $\sensing: \mathbb{R}^{\dimension} \rightarrow \mathbb{R}^{\numsam} $ is a linear operator satisfying rank-RIP and sparse-RIP with constants $\delta_{4\rank}(\sensing) \leq 0.09$ and $\delta_{4\sparsity}(\sensing) \leq 0.095$, respectively, and satisfies jointly the low rank- and sparse-RIP with constant $\delta_{3\rank + 3\sparsity}(\sensing) \leq 0.095$, we observe that the eigenvalues of $\widehat{\boldsymbol{\Delta}}$ are distinct and real and satisfy $|\lambda_j(\widehat{\boldsymbol{\Delta}})| < 1, ~\forall j$. Furthermore, $| \mathbb{I} - \widehat{\boldsymbol{\Delta}} | \neq 0 $. To complete the proof, we use the following Theorem from \cite{DDS} --- the proof is omitted:
\begin{theorem}[Necessary and Sufficient Conditions for Global Stability: Distinct Real Eigenvalues]
Consider the system $\mathbf{w}(i+1) = \widehat{\boldsymbol{\Delta}} \mathbf{w}(i) + \mathbf{B} $ where $\mathbf{w}(0)$ is given. We assume that $|\mathbb{I} - \widehat{\boldsymbol{\Delta}}| \neq 0$ and $\widehat{\boldsymbol{\Delta}}$ has distinct real eigenvalues. Then:
\begin{itemize}
\item The steady-state equilibrium $ \widetilde{\mathbf{w}} = [\mathbb{I} - \widehat{\boldsymbol{\Delta}}]^{-1}\mathbf{B} $ is globally stable if and only if $|\lambda_j(\widehat{\boldsymbol{\Delta}})| < 1, ~\forall j$.
\item $\lim_{i \rightarrow \infty} \mathbf{w}(i) = \widetilde{\mathbf{w}} $ if and only if $|\lambda_j(\widehat{\boldsymbol{\Delta}})| < 1, ~\forall j$.
\end{itemize}
\end{theorem} 

In our simple case, we consider $\mathbf{B} := 0 $. Thus, the steady-state equilibrium in (\ref{recursive}) satisfies $ \widetilde{\mathbf{w}} = \mathbf{0} $. Then, we conclude $\lim_{i \rightarrow \infty} \mathbf{w}(i) = \mathbf{0} $ and, thus:
\begin{align}
\vectornormbig{\mathbf{L}_i - \mathbf{L}^{\ast}}_F \rightarrow 0 ~~~~\text{and}~~~~ \vectornormbig{\mathbf{M}_i - \mathbf{M}^{\ast}}_F \rightarrow 0,
\end{align} as $i \rightarrow \infty$.

\bibliographystyle{unsrt}
\bibliography{recipes}

\end{document}